\documentclass[aps, prx, reprint, longbibliography, nofootinbib,superscriptaddress, floatfix]{revtex4-1}
\usepackage{slashed}
\usepackage{verbatim}
\usepackage[T1]{fontenc}
\usepackage{mathbbol}
\usepackage[dvipsnames]{xcolor}
\usepackage{orcidlink}

\usepackage[normalem]{ulem}
\usepackage[english]{babel}
\usepackage{lipsum}
\usepackage{physics}
\usepackage{dcolumn}
\usepackage{multirow}
\usepackage{tensor}
\usepackage{comment}
\usepackage{placeins}
\usepackage{graphicx,color,overpic,mathtools}
\usepackage{amsthm,amsmath,amssymb,mathrsfs}
\usepackage{braket,bm,bbm,setspace}
\usepackage{booktabs}
\usepackage{cancel}
\usepackage{float}
\usepackage{xargs}
\definecolor{myred}{RGB}{179, 27, 27}
\usepackage{hyperref}
\hypersetup{
    colorlinks=true,
    linkcolor=myred, 
    citecolor=myred, 
    urlcolor=myred  
 }
 
\def\be{\begin{equation}}
\def\ee{\end{equation}}
\def\bea{\begin{eqnarray}}
\def\eea{\end{eqnarray}}
\def\beq{\begin{eqnarray}}
\def\eeq{\end{eqnarray}}
\begin{document}

\title{Ringdown and lensing of triple systems}

\author{Vitor Cardoso\orcidlink{0000-0003-0553-0433}}
\affiliation{Center of Gravity, Niels Bohr Institute, Blegdamsvej 17, 2100 Copenhagen, Denmark}
\affiliation{CENTRA, Departamento de F\'{\i}sica, Instituto Superior T\'ecnico -- IST, Universidade de Lisboa -- UL,
Avenida Rovisco Pais 1, 1049 Lisboa, Portugal}

\author{Giuseppe Ficarra\orcidlink{0000-0003-4668-8587}}
\affiliation{Dipartimento di Fisica, Universit\'a della Calabria, Arcavacata di Rende (Cosenza), 87036, Italy}

\author{Jaime Redondo-Yuste\orcidlink{0000-0003-3697-0319}}
\affiliation{Center of Gravity, Niels Bohr Institute, Blegdamsvej 17, 2100 Copenhagen, Denmark}
\affiliation{William H. Miller III Department of Physics and Astronomy, Johns Hopkins University,
3400 North Charles Street, Baltimore, Maryland, 21218, USA}

\author{João Sieiro dos Santos\orcidlink{0000-0003-1921-8758}}
\affiliation{CENTRA, Departamento de F\'{\i}sica, Instituto Superior T\'ecnico -- IST, Universidade de Lisboa -- UL,
Avenida Rovisco Pais 1, 1049 Lisboa, Portugal}

\begin{abstract}
Triple systems have progressively been recognized as ubiquitous in our universe and provide a good testing ground for wave generation and propagation in nontrivial environments. We study the dynamics of triple systems in a fully nonlinear setting. In particular, we analyze numerical relativity simulations of head-on collisions of black holes in the presence of a companion. We show evidence for Doppler and gravitational redshift in the ringdown, and clear signs of amplification by lensing. In certain cases, we also show the appearance of a second image, with hints of resonant mode excitation. Our results pave the way for the understanding of mergers in the vicinity of massive companions.
Even in extreme setups we do not find collapse to black holes from lensed gravitational radiation.
\end{abstract}
\maketitle
%\tableofcontents
%%%%%%%%%%%%%%%%%%%%%%%%
\section{Introduction}
%%%%%%%%%%%%%%%%%%%%%%%%
%
Isolated, equilibrium solutions of Einstein's theory are well studied. In vacuum, the most general stationary, regular solution which is also asymptotically flat belongs to the Kerr family of black holes (BHs)~\cite{Bekenstein:1996pn,Carter:1997im,Chrusciel:2012jk,Robinson:1975bv}. This powerful result, together with the impossibility of massive objects to guard against total gravitational collapse, promotes Kerr BHs to ideal laboratories to test General Relativity: they exist, and they are simple~\cite{Cardoso:2019rvt,Cardoso:2025npr}. A sizable fraction of BHs are {\it not} in isolation, but rather evolve together with a companion in binary systems~\cite{Shakura:1972te,LIGOScientific:2018jsj}. In the late stages of the life of such a binary, the dynamics are driven by gravitational radiation. The relevance of this process for gravitational-wave astronomy and our knowledge of motion in General Relativity spurred a fantastic effort to understand the details of the two-body problem in curved spacetime (see, e.g., Refs.~\cite{Blanchet:2013haa,Bern:2019nnu,Buonanno:2006ui,Berti:2007fi,Scheel:2025jct,Berti:2025hly}). 

It is clear that the degree of complexity of bound systems increases almost indefinitely as the number of constituents grows. Triple systems play a special role in this catalog. A growing body of work suggests they may abound in active galactic nuclei and globular clusters, and play a crucial role in the hardening of two-body systems~\cite{Bartos:2016dgn,Stone:2016wzz,Peng:2024wqf,Sberna:2022qbn,Dittmann:2023sha,Yang:2024tje,LIGOScientific:2025brd, Phukon:2025cky,Li:2025fnf, Stegmann:2025zkb,Chen:2018axp,Han:2018hby,Yin:2024nyz,Jiang:2024mdl,Zevin:2018kzq, Tagawa:2019osr,Martinez:2020lzt,Hoang:2017fvh, Bellovary:2015ifg,Antonini:2015zsa, Antonini:2012ad}. In addition, they are the simplest gravity-governed system which comprises different combinations of emitters: each pair emits gravitational waves, leading to clear lensing signatures~\cite{Kubota:2024zkv, Pijnenburg:2024btj,Chan:2025wgz,DOrazio:2019fbq,Oancea:2022szu,Oancea:2023hgu,Ubach:2025oxr,Saketh:2025cwf}, Doppler shifts~\cite{Cisneros:2012sk,Bonvin:2016qxr,Inayoshi:2017hgw,Robson:2018svj,Tamanini:2019usx,Meiron:2016ipr, Wong:2019hsq,Randall:2018lnh,Yan:2023pyo,Zwick:2025wkt,Takatsy:2025bfk, Fang:2025vnv}, beaming~\cite{Torres-Orjuela:2018ejx,Torres-Orjuela:2020cly,Torres-Orjuela:2020oxq,Bonvin:2022mkw,Cusin:2024git}, resonances~\cite{Cardoso:2021vjq,Santos:2026lzq,Cocco:2025adu,Cocco:2025udb,Camilloni:2023xvf}, and more. They represent a superb setup to test a number of features of highly relativistic, nontrivial systems. 

As a pair in the triple coalesces, it emits a burst of radiation which is dominated by the characteristic modes of the newly-formed BH~\cite{Berti:2009kk,Berti:2025hly}. This radiation will be lensed by the companion, and possibly excite its own modes. In recent years, the BH spectroscopy program has proved capable of providing clean, inspiral-merger independent, analysis of gravitational wave events, and doing so robustly testing the Kerr hypothesis~\cite{Berti:2025hly}. The program hinges on the idea that the final, ringdown, stage of a binary BH coalescence, can be well described by a sum of quasinormal modes (QNMs), whose frequencies depend only on the properties of the remnant. However, an astrophysical environment could modify the relation between remnant and QNM frequencies, breaking the assumptions of BH spectroscopy. It was shown in Ref.~\cite{Spieksma:2024voy} that this is not the case for dilute, halo-like environments. The same may not hold for a ringdown in the presence of a third compact object, a case for which results are currently limited to eikonal limit analyses~\cite{Cardoso:2021qqu, Katagiri:2026gkz,Cocco:2026lkr,Camilloni:2023rra}.

In a recent work~\cite{Santos:2025ass}, some of the authors developed, from first principles, a perturbative model for compact triples with a scale hierarchy in the strong field (known as ``b-EMRIs'' in the literature~\cite{Chen:2018axp,Han:2018hby}). Strong evidence and detailed studies of all the phenomenology above was put forward, but the linear and non-evolving nature of the model means it fails to capture a number of interesting effects. Strong nonlinear phenomena might lead to qualitatively new effects~\cite{Ma:2025rnv,Cardoso:2026llh}. Moreover, the extrapolation of a point-particle approximation to equal masses may not scale as smoothly as in the two-body problem~\cite{Kuchler:2024esj}.

\begin{figure}[t]
    \centering
    \usetikzlibrary{arrows.meta,calc}

\definecolor{myred}{HTML}{C40000}
\definecolor{myblue}{HTML}{0053A1}

\definecolor{mylgray}{HTML}{D3D3D3}
\definecolor{mydgray}{HTML}{808080}

\begin{tikzpicture}[
    >=Latex,
    font=\normalsize,
    axis/.style={->, thin},
    mergearrow/.style={->, line width=0.8pt},
    mass/.style={circle, inner sep=0pt, minimum size=8pt},
    massR/.style={mass, fill=myred},
    massB/.style={mass, fill=myblue},
    massK/.style={mass, fill=black},
    remnant/.style={circle, draw=black, fill=none, minimum size=9pt, inner sep=0pt}
]

% ------------------------------------------------------------
% Coordinates
% ------------------------------------------------------------
\coordinate (O)  at (0,0);
\coordinate (M1) at (-3.2,0);

\def\d{0.85} % half the side length of the square

% center of the merger
\coordinate (C) at (2.7,0);

% red = aligned configuration
\coordinate (R2) at ($(C)+(-\d,0)$);
\coordinate (R3) at ($(C)+(\d,0)$);

% blue = non-aligned configuration
\coordinate (B2) at ($(C)+(0,-\d)$);
\coordinate (B3) at ($(C)+(0,\d)$);

% ------------------------------------------------------------
% Axes
% ------------------------------------------------------------
\draw[axis] (-4.2,0) -- (4.4,0) node[right] {$x$};
\draw[axis] (0,-1.2) -- (0,1.2) node[above] {$z$};

% ------------------------------------------------------------
% Common mass
% ------------------------------------------------------------
\node[massK] at (M1) {};
\node[above=4pt] at (M1) {$m_1$};

% ------------------------------------------------------------
% Red aligned configuration
% ------------------------------------------------------------
\node[massR] at (R2) {};
\node[massR] at (R3) {};

% ------------------------------------------------------------
% Blue vertical configuration
% ------------------------------------------------------------
\node[massB] at (B2) {};
\node[massB] at (B3) {};

% ------------------------------------------------------------
% Remnant at the center
% ------------------------------------------------------------
\node[remnant] at (C) {};

% ------------------------------------------------------------
% Labels
% ------------------------------------------------------------
\node[text=black, below=18pt, right=-3pt] at (R2) {$m_2$};
\node[text=black, above=18pt, left=-3pt] at (R3) {$m_3$};

% ------------------------------------------------------------
% Merger arrows
% ------------------------------------------------------------
% red arrows (horizontal, inward)
\draw[mergearrow, myred] ($(R2)!0.30!(C)$) -- ($(R2)!0.70!(C)$);
\draw[mergearrow, myred] ($(R3)!0.30!(C)$) -- ($(R3)!0.70!(C)$);

% blue arrows (vertical, inward)
\draw[mergearrow, myblue] ($(B2)!0.30!(C)$) -- ($(B2)!0.70!(C)$);
\draw[mergearrow, myblue] ($(B3)!0.30!(C)$) -- ($(B3)!0.70!(C)$);

\end{tikzpicture}
    \caption{A cartoon of the setup studied in this work and listed in Tab.~\ref{tab:nr_quantities}. Three BHs, of masses $m_1, m_2, m_3$ are initially at rest. Initial data of type AE, AU has all BHs collinear (red), while in in UE, UU $m_2$ and $m_3$ are perpendicular to the $x$ axis (blue). The masses $m_2,\,m_3$ merge first (arrows) and form a remnant on the $x$ axis (black circle). The radiation emitted in this merger is scattered by mass $m_1$. Finally, the end product of collision between $m_2,\,m_3$ merges with $m_1$, producing a final non-spinning, but possibly moving BH.}
    \label{fig:collinear_sketch}
\end{figure}
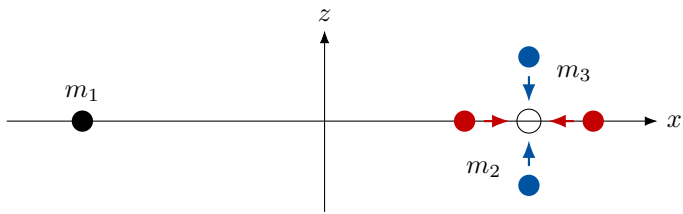
The relevance of this problem stimulated several studies evolving the full set of Einstein equations to understand systems featuring BH triplets or more~\cite{Bai:2011za,Campanelli:2007ea,Lousto:2007rj,Imbrogno:2021xrh,Ficarra:2023zjc,Ficarra:2024jen,Bamber:2025gxj,Heinze:2025usf,Wu:2026hth}. These works investigated the kinematics and time variability of coalescence events, and provided a qualitative assessment of the complexity displayed by gravitational waves generated from such systems.
Here, we use fully nonlinear numerical relativity simulations of three body systems in simple collinear initial data setups to investigate the following questions: \textit{(i)} how is the ringdown affected by the presence of a third compact object? \textit{(ii)} can a BH significantly focus radiation scattering off it? and \textit{(iii)} can nonlinearities be enhanced in triples, where the scale invariance of binaries no longer holds?

Our work is structured as follows: In Sec.~\ref{sec:framework} we present the physical setup, main analytical results, and  important aspects of the numerical simulations. We present the results of the simulations in Sec.~\ref{sec:results} and test them for evidence of frequency shifts, lensing/scattering, resonant mode excitation, and non-linear effects. We conclude in Sec.~\ref{sec:discussion}, summarizing and laying out future work.

%%%%%%%%%%%%%%%%%%%%%%%%%%%%%%%%%%%%%%%%%%%%%%%
\section{Framework} \label{sec:framework}
%%%%%%%%%%%%%%%%%%%%%%%%%%%%%%%%%%%%%%%%%%%%%%%
%%%%%%%%%%%%%%%%%%%%%%%%%%%%%%%%%%%%%%%%%%%%%%%
\subsection{The setup and Newtonian estimates for dynamical timescales}
%%%%%%%%%%%%%%%%%%%%%%%%%%%%%%%%%%%%%%%%%%%%%%%
%
\begin{table*}[ht!]
    \begin{tabular}{c|c|c|c|c|c|c|c|c|c|c|c|}
        ID      & $m_2$  & $m_3$  & $m_1$  &$(x_1,x_2,x_3)$& $M_1$  & $M_2$ & $t_{M_1}$ & $t_{M_2}$      & $10^{3}v_{\rm M1}$ & $\Delta v_{\rm M1}/v_{\rm M1}$ & $R_{\rm H23}/ d_{23}$ \\
    \hline
    
%A
{\bf ID AE} & $1/3$ & $1/3$ & $1/3$ &(-20, 15, 25)  & 0.6555 & 0.9830 & 58.0 (43.0) & 312.2 (281) & 9.1 & 0.78 & 5.8 \\
%
%{\bf ID B:} & 0.3333 & 0.3333 & 0.3333 &(-40, 35, 45)  & 0.6556 & 0.9856 & 56.7 (43.0) & 842.0 (795) & 2.2 & 0.77 & 11 \\
%C
{\bf ID AU} & $1/4$ & $1/4$ & $1/2$ &(-20, 15, 25)  & 0.4936 & 0.9870 & 62.1 (49.7) & 310.7 (281) & 16 & 0.73 & 5.0 \\    
%
%{\bf ID D:} & 0.2500 & 0.2500 & 0.5000 &(-40, 35, 45)  & 0.4936 & 0.9899 & 61.3 (49.7) & 841.1 (795) & 3.9 & 0.71 & 10 \\
%
%{\bf ID E$^*$:} & 0.2500 & 0.2500 & 0.5000 &(-20, 35, 35)  & 0.4937 & 0.9886 & 60.8 (49.7) & 490.8 (453) & 8.2 & 0.72 & 6.9 \\
%
%F
{\bf ID UU} & $1/4$ & $1/4$ & $1/2$ &(-20, 20, 20)  & 0.4937 & 0.9871 & 61.1 (49.7) & 313.3 (281) & 16 & 0.73 & 5.0 \\ 
%G
{\bf ID UE} & $1/3$ & $1/3$ & $1/3$ &(-20, 20, 20)  & 0.6557 & 0.9831 & 55.2 (43.0) & 313.2 (281) &9.1 & 0.78 & 5.8 \\
%RA
{\bf ID AB} & $1/3$ & $1/3$ & $0$ &(-, 15, 25)  & 0.6555 & - & 54.3 (43.0) & - (-) & - & - & - \\
%S
{\bf ID UB} & $1/4$ & $1/4$ & $0$ &(-, 20, 20)  & 0.4936 & - & 59.7 (49.7) & - (-) & - & - & -
    \end{tabular}
    \caption{Initial data setup and main results from the nonlinear evolution. Here, $m_1$, $m_2$ and $m_3$ describe the triplet initial masses as measured from the apparent horizon finder, while $M_1$ and $M_2$ are the BH masses from the first and second merger, respectively. The initial data consists of BHs aligned (A) in a collinear configuration, or un-aligned (U) at right angles. The masses are either all equal (E) or unequal (U). For aligned runs, BHs are placed at $z=0$, for unaligned runs they sit at $(z_1, z_2, z_3)=(0, -5, 5)$. The reference binary (B) runs have $m_1=0$.
    We also list the merger times $t_{M_1}$ and $t_{M_2}$ defined as the time of formation of the common apparent horizon.
    In parenthesis we list the Newtonian prediction \eqref{eq:merger_time}.
    All quantities are in units of the total mass $M$ of the system. We also show $v_{\rm M1}$, the velocity of the remnant M1 after merger, as estimated at Newtonian level; $\Delta v_{\rm M1}$, the fractional change in center of mass velocity of remnant M1 during ringdown, Eq.~\eqref{eq:ringdown_time}; finally, $R_{\rm H23}/ d_{23}$ is a measure of the hardness of the system $m_2,\,m_3$ at $t=0$ (Eq.~\eqref{eq:Hills})-- all setups being safe against tidal disruption. 
    }
    \label{tab:nr_quantities}
\end{table*}
%
%Our strategy is to solve exactly the constraints for 3 BHs initially at rest, with one much farther away than the other two. \jr{this sentence is very strange to start this section. I propose, simply: "
We consider a simple configuration of three nonspinning BHs close to alignment. One of the BHs is much farther away than the other two, {and the BHs start from rest, allowing us to solve exactly the constraints at the level of the initial data. 
%\gf{We can solve the constraints exactly also because no initial spin or momenta} 
The setup is illustrated in Fig.~\ref{fig:collinear_sketch}, either collinear or at right angles: three masses ($m_1, m_2, m_3$), starting from rest in the x-axis at locations ($x_1, x_2, x_3$), and in the $z$-axis at locations ($0, Z,-Z)$. For the collinear configurations, $Z=0$, while for the configurations at right angles, $x_2=x_3$ and $Z> 0$. The closest two BHs ($m_2$ and $m_3$) merge and the remnant relaxes, ringing down in its characteristic modes. Eventually, this remnant would itself merge with the companion farther away, producing a second ringdown stage.

The ringdown of BHs in a strong tidal field is not a setup where we can expect to have great analytical control. For this reason, it is important to isolate the contributions arising from different phenomena as much as possible. Some effects under clean, analytic control include the motion of the center of mass of the remnant as it rings, which will imprint a Doppler shift, and the gravitational redshift due to the curvature generated by the third body. One has good control over both of these frequency shifts, if they are constant. If instead they are varying rapidly over the ringdown, their effect on the signal may be hard to predict and isolate. Thus, an estimate of timescales is important.

We are interested in hierarchical setups with $|x_1-x_{2,3}| \gg |x_2-x_3|$, so we will find estimates separating the evolution into its two distinct stages: merger M1, when masses $m_2$ and $m_3$ coalesce; and merger M2 when the remnant of M1 merges with mass $m_1$. With this in mind, the collision time of two point masses $m_1$, $m_2$, in Newtonian gravity starting from rest at total separation $d$ is
\be
\left(t_{\rm M}^{\rm N}\right)^2=\frac{\pi^2d^3}{2^3 (m_1+m_2)}\,. \label{eq:merger_time}
\ee
The other relevant timescale is that of the ringdown, which, for a BH of mass $M$ is~\cite{Berti:2009kk,Berti:2025hly}
\be
t _{\rm RD} \approx \mathfrak{I}[M \omega_{20}]^{-1} \approx 22 M \,,\label{eq:ringdown_time}
\ee
where $\omega_{20}$ is the dominant quadrupolar ringdown frequency~\cite{Berti:2009kk,Berti:2025hly}, see also Eq.~\eqref{eq:ringdown_model} below. 
We can use these results to obtain estimates for the velocity of the center of mass of the remnant of M1 at merger $v_{\rm M1}$ as well as the variation in its velocity during the ringdown $\Delta v_{\rm M1}$.

%%
%\beq
%{\bf ID A:}\, (m_1, m_2, m_3)=(1,1,1),\, (x_1, x_2, x_3)=(-20, 15, 25)\,,\nonumber\\
%%
%{\bf ID B:}\, (m_1, m_2, m_3)=(1,1,1),\, (x_1, x_2, x_3)=(-40, 35, 45)\,,\nonumber \\
%%
%{\bf ID C:}\, (m_1, m_2, m_3)=(2,1,1),\, (x_1, x_2, x_3)=(-20, 15, 25)\,,\nonumber \\
%%
%{\bf ID D:}\, (m_1, m_2, m_3)=(2,1,1),\, (x_1, x_2, x_3)=(-40, 35, 45)\,.\nonumber\\
%%
%{\bf ID E:}\, (m_1, m_2, m_3)=(2,1,1),\, (x_1, x_2, x_3)=(-20, 20, 20),\,(z_1, z_2, z_3)=(0, -5, 5)\,.\nonumber
%\eeq
%%

Another important quantity to have under control is the comparison between the tidal forces that the system $(m_2,m_3)$ is subject to and its self-energy. This is encapsulated by the Hills criterion~\cite{Hills_1988}: the M1 system will not be tidally disrupted if
\begin{equation}
R_{\rm H23}/d_{23} = ((m_2+m_3)/m_1)^{1/3} d_0/d_{23} > 1 \, ,
\label{eq:Hills}
\end{equation}
where $R_{\rm H23}$ is the Hills radius of the $(m_2,m_3)$ binary in the field of $m_1$ and $d_{23}$ is the distance between $m_2$ and $m_3$, and $d_{0}$ is the distance between the center of mass of the former and $m_1$. 

Keeping these in mind, we have settled on the six configurations summarized in Table~\ref{tab:nr_quantities}, where we also show
the characteristic timescales and Hills criteria. %\sout{The initial data (ID) includes two reference runs where the companion does not exist, $m_1=0$ for ID AB and UB, details on numerical implementation are given below. Because radiation emitted along the axis of collision vanishes, and because we are interested in understanding lensing and resonance, we also evolved initial data un-aligned, where the first merger takes place orthogonally to the axis joining $m_1$ and the geometric center of $m_2,\,m_3$.} 
The initial data (ID) are labeled according to their geometric orientation and the BH masses. The first four runs have three BHs which can be aligned (ID A...; red in Fig.~\ref{fig:collinear_sketch}) or un-aligned (ID U...; blue in Fig.~\ref{fig:collinear_sketch}). In head-on mergers, no radiation is emitted along the axis of collision, so ID U... runs will have more radiation interacting with $m_1$. The runs can also have all BHs with equal mass (ID ...E), or unequal masses, with a $2:1:1$ ratio (ID ...U). The latter are interesting because $M_1 \approx m_1$, which may lead to resonant excitation of modes.
We also include two reference runs where the companion does not exist, $m_1=0$, for ID AB (aligned) and UB (unaligned). Details on the numerical implementation of the ID are given below. 
\subsection{Doppler and gravitational frequency shifts\label{subsec:Doppler}}
%%%%%%%%%%%%%%%%%%%%%%%%%%%%%%%%%%%%%%%%%%%%%%%
%
When the first merger M1 takes place, the remnant is moving with respect to the center of mass. Thus, characteristic waves will be Doppler shifted with respect to observers at rest (with respect to the center of mass). In flat space, this motion causes the observed frequency $\omega_{\rm obs}$, to differ from the emitted frequency $\omega_{\rm em}$, as
\begin{equation}
\frac{\omega_{\rm obs}}{\omega_{\rm em}} = \left[\gamma (1 - v \cos \theta)\right]^{-1} \equiv f_{\rm Doppler} \, ,
\end{equation}
where $v$ is the relative velocity between BH and observer and $\theta$ the respective angle to the line of sight. Another important effect will be the gravitational redshift, which is the result of time passing more slowly in strong gravitational fields. Since the first merger occurs in the field of mass $m_1$ in Fig.~\ref{fig:collinear_sketch}, the ringdown waves will also be affected by gravitational redshift. Using a weak field and slow motion approximation, we estimate this redshift to be 
\begin{equation}
\frac{\omega_{\rm obs}}{\omega_{\rm em}} = \sqrt{\frac{1-2 m_1/r_{\rm obs}}{1-2 m_1/r_{\rm em}}} \equiv f_{\rm Grav} \, . 
\end{equation}
Note that, because the first merger occurs relatively far from $m_1$, both factors $f$ are very close to unity (up to about 1\%). For this reason, to get the total frequency shift from the combined effects we can take
\begin{equation}
\frac{\omega_{\rm obs}}{\omega_{\rm em}}  =  f_{\rm Doppler}  f_{\rm Grav} \, .\label{eq:shift}
\end{equation}
%

%%%%%%%%%%%%%%%%%%%%%%%%%%%%%%%%%%%%%%%%%%%%%%%
\subsection{Numerical Implementation}
%%%%%%%%%%%%%%%%%%%%%%%%%%%%%%%%%%%%%%%%%%%%%%%
In order to simulate fully relativistic BH triples, we employ the \textsc{Einstein Toolkit}~\cite{roland_haas_2024_14193969,EinsteinToolkitWeb}, an open-source framework for numerical relativity and computational astrophysics built upon the \textsc{Cactus} computational toolkit~\cite{Goodale:2002a,Cactuscode:web}, together with the \textsc{Carpet} mesh refinement driver~\cite{Schnetter:2003rb,CarpetCode:web}. The physics modules are supplied by the open-source \textsc{Canuda} code for numerical relativity in fundamental physics~\cite{witek_helvi_2023_7791842}.

We evolve the metric equations using the BSSN formulation~\cite{Baumgarte:1998te,Shibata:1995we} as described in \cite{Cheng:2025wac}, in conjunction with the moving puncture gauge~\cite{Campanelli:2005dd,Baker:2005vv,Alcubierre:2002kk}, supplemented by the slow-start lapse condition~\cite{Etienne:2024ncu}. Numerical integration is carried out via the method of lines, employing fourth-order finite differences for spatial derivatives and a fourth-order Runge-Kutta time integrator, along with a fifth-order Kreiss-Oliger dissipation operator with coefficient $\epsilon_{\rm diss} = 0.2$ and Sommerfeld radiative boundary conditions ~\cite{Alcubierre:2002kk}.

Gravitational radiation is extracted at selected points in the domain via the computation of the Newman-Penrose scalar $\Psi_{4}$, following the implementation described in~\cite{Cheng:2025wac}. Local properties of BHs are computed using the \textsc{QuasiLocalMeasures} thorn~\cite{Dreyer:2002mx}, while their apparent horizons are located with the \textsc{AHFinderDirect} thorn~\cite{Thornburg:2003sf}.

Since we consider non-spinning BHs initially at rest, we adopt Brill-Lindquist-type initial data~\cite{Brill:1963yv,Lindquist1963}. The normalization is chosen so that the total mass of the spacetime $M$ equals the sum of the individual BH masses $m_i$ as measured by the apparent horizon finder.

The simulation domain is a three-dimensional Cartesian grid with an extent of $\sim 512M$ in each dimension. We employ 10 levels of moving, box-in-box style mesh refinement, such that the finest grid spacing centered around each BH is $\sim M/185$. The resolution on the outermost refinement level is $\Delta x_{\rm c} \simeq 2.77M$, while the resolution in the extraction zone is $\Delta x_{\rm ext} \simeq 0.69M$. For practical purposes, we set $M=1$ in the simulations. We provide convergence studies for the numerical simulations considered in this work in Appendix~\ref{app:convergence}.

%%%%%%%%%%%%%%%%%%%%%%%%%%%%%%%%%%%%%%%%%%%%%%%
\section{Numerical Results} \label{sec:results}
%%%%%%%%%%%%%%%%%%%%%%%%%%%%%%%%%%%%%%%%%%%%%%%
%%%%%%%%%%%%%%%%%%%%%%%%%%%%%%%%%%%%%%%%%%%%%%%
\subsection{Qualitative features and timescales}
%%%%%%%%%%%%%%%%%%%%%%%%%%%%%%%%%%%%%%%%%%%%%%%
%
\begin{figure*}
    \centering 
    \includegraphics[width=0.49\linewidth]{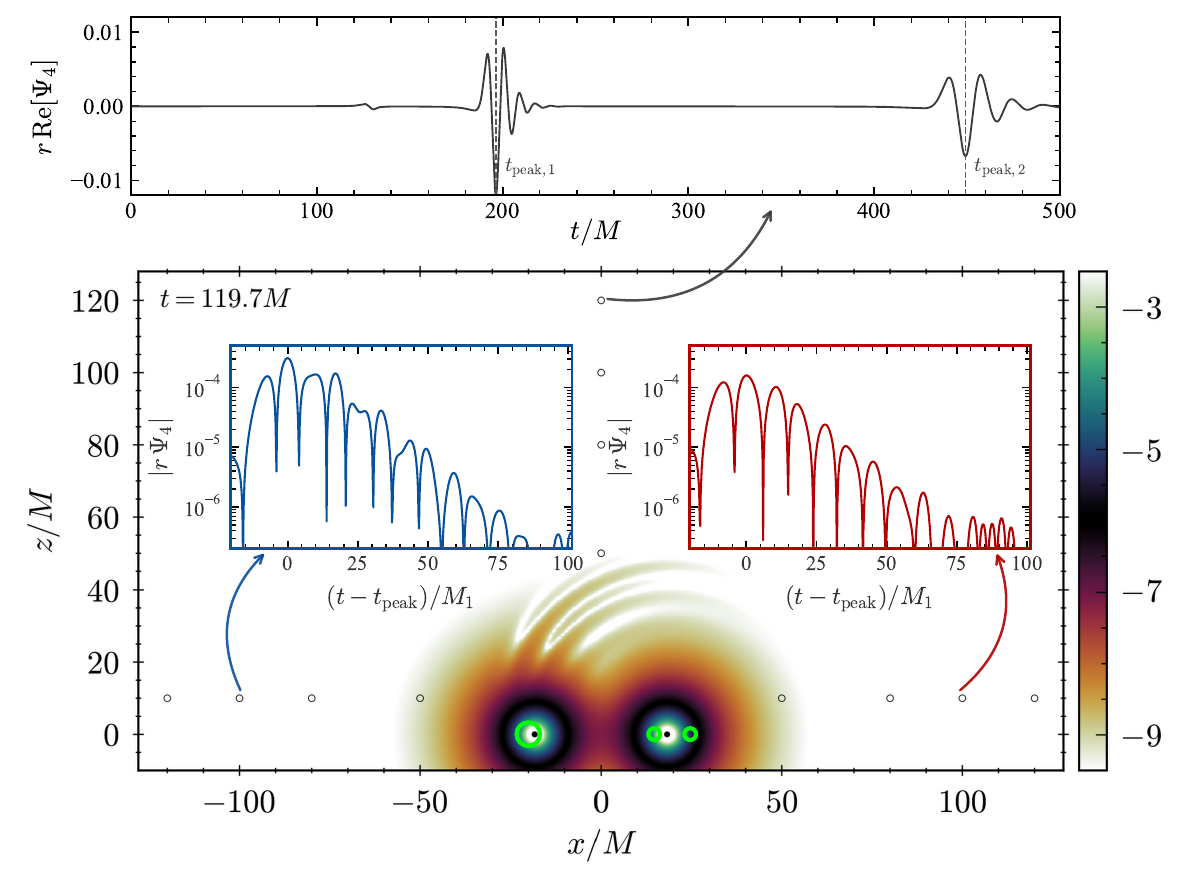}
    \includegraphics[width=0.49\linewidth]{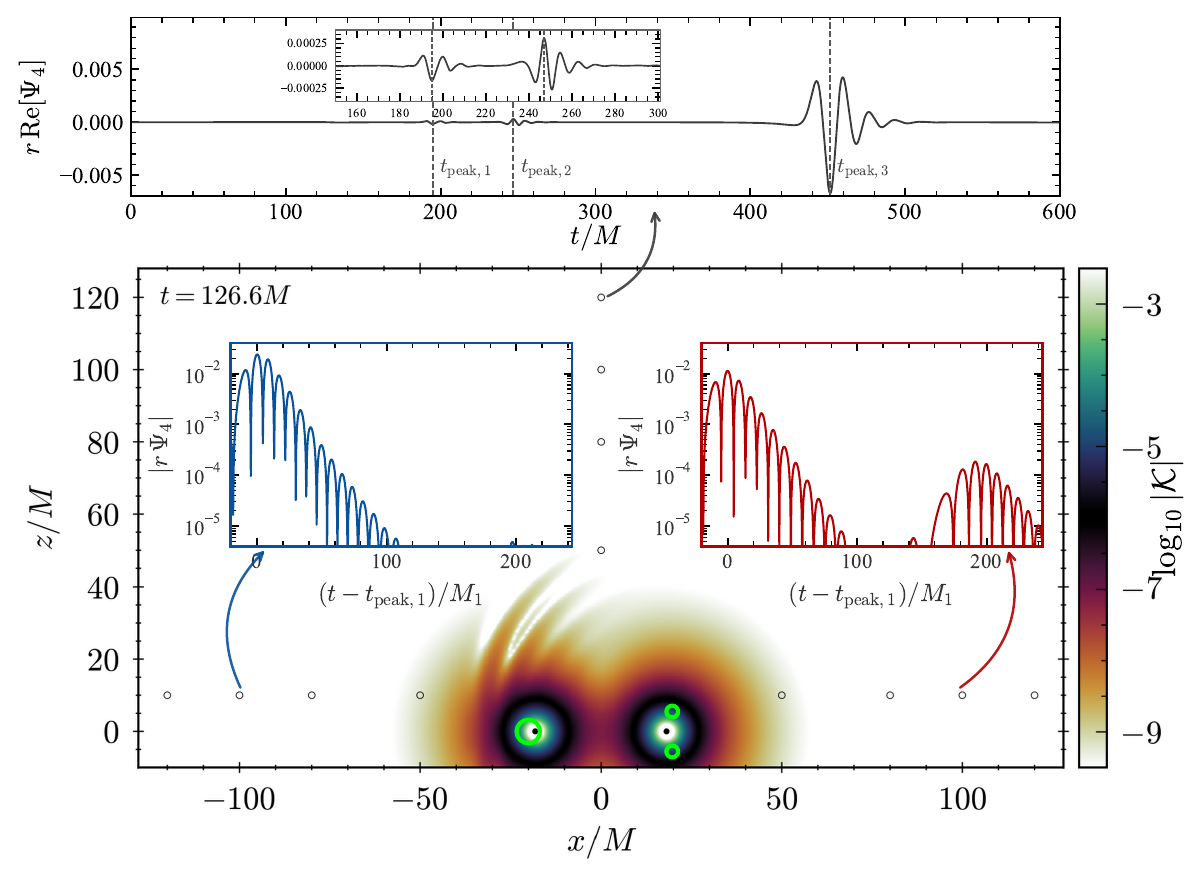}
    \caption{Summary of two of the configurations considered in this work, corresponding to initial data ID AU (\textbf{left}) and ID UU (\textbf{right}). The bottom panel shows a snapshot of the Kretschmann scalar, $\log_{10} |\mathcal K|$, directly extracted from the simulations. The empty circles indicate some of the observing points we consider. The green circles denote the initial location of the three BHs, with size representing their relative masses (see also Fig.~\ref{fig:collinear_sketch} and Tab.~\ref{tab:nr_quantities}). The inset panels show the gravitational waves extracted at the indicated observing points ($x=\pm 100, z=10$) following the first merger. The top panel shows the whole gravitational-wave strain during the two consecutive mergers, extracted at $x=0,z=120$. Green circles mark initial location of horizons. Videos of the numerical simulations can be found in~\cite{YoutubeCanuda, web:CoG}.
    }
    \label{fig:summary}
\end{figure*}
Our numerical results are summarized in Table~\ref{tab:nr_quantities} and Fig.~\ref{fig:summary}.
As seen in Table~\ref{tab:nr_quantities}, a Newtonian estimate for the merger time is in good agreement with our nonlinear simulations, and allows us to estimate the relative velocity of merger remnant M1 when the first ringdown starts; in particular we expect Doppler effects of order 1\%.

Consider, then, the first merger (the second takes place in isolation and we will never consider it in any detail). When M1 takes place, gravitational waves are emitted across the celestial sphere. Some of that radiation travels unimpeded to the observer, but other interacts first with mass $m_1$ before reaching the observer. Figure~\ref{fig:summary} shows two interesting features, which arise as a consequence of such interaction. The first is a clear morphology change of the signal due to lensing or scattering imprint in the waveform. In particular, when extracting close to the axis of collision in ID AE (see left panel) the gravitational-wave signal to the right, travels unimpeded and we see a clear ringdown waveform. However, for an observer on the negative x-axis, the waves had to interact with mass $m_1$: they are lensed or scattered from it, they interfere, and this causes a visible distortion in the signal. This is not seen in ID UU because there the signal there is strong on the axis, which is now where extraction is done (rather than slightly off it). This increased symmetry of the lensing setup implies the signal is less distorted. 
%
%\jr{in Fig.~\ref{fig:summary} i am extracting slightly off axis also on the right panel. i think the reason why the distorsions are not so evident is simply that the amplitude of the ``fist image'' is way louder, whereas for AE the first image is almost zero, so we see a large overlap between images...}

The second striking feature is that non collinear configurations, e.g., ID UU and UE, give rise to a double ringdown structure following the first merger, for certain observation points. This is a second image -- in the lensing language -- or an ``echo'' of the first ringdown, that instead of propagating directly from the merger site to the observer's location, is deflected in the strong field of the BH companion.
The amplitudes of the first and second images are different: when the observer is on the $z$ axis (see the top right panel of Fig.~\ref{fig:summary}), it sees the merger almost edge-on, and hence the first image is suppressed (no radiation is emitted along the axis of collision). The second image, which in the geometric optics limit can be thought of as taking a half orbit around the BH companion, is a reflection of the radiation that is emitted face-on, and hence has higher amplitude. When the observer is on the \emph{right} side relative to the merger (e.g., the red inset in the right panel of Fig.~\ref{fig:summary}), the first image is seen face-on, and has a large amplitude. The second image is more delayed and more suppressed, as it performs a full orbit around the BH. 

It is also apparent (blue and red ringdown insets in Fig.~\ref{fig:summary}) that the signal as seen to the left of mass $m_1$ (blue) is magnified, compared to the amplitude of the signal in the right (red). Close inspection reveals this magnification factor to be of order $\sim 2$. We will examine this feature more closely in Section~\ref{sec:lensing}, but we can already anticipate that this is an expected feature of gravitational-wave lensing. 
%Another smoking gun of lensing effects in the signal is the different morphology of the ringdown as observed to the left and the right for ID C (left panel of Fig.~\ref{fig:summary}). This hints at possible mode-beating effects. We investigate the mode content of these ringdown signals in the upcoming sections. 

Thus, this is a very rich system, with Doppler and gravitational shifts, lensing/scattering, etc. For some parameters, the ringdown from merger M1 might even excite resonantly the companion BH. In the following we quantify some of the most interesting features in the waveform.

%%%%%%%%%%%%%%%%%%%%%%%%%%%%%%%%%%%%%%%%%%%%%%%
\subsection{First image} \label{sec:direct}
%%%%%%%%%%%%%%%%%%%%%%%%%%%%%%%%%%%%%%%%%%%%%%%%

\begin{figure*}[t]
    \centering
    \includegraphics[width=\linewidth]{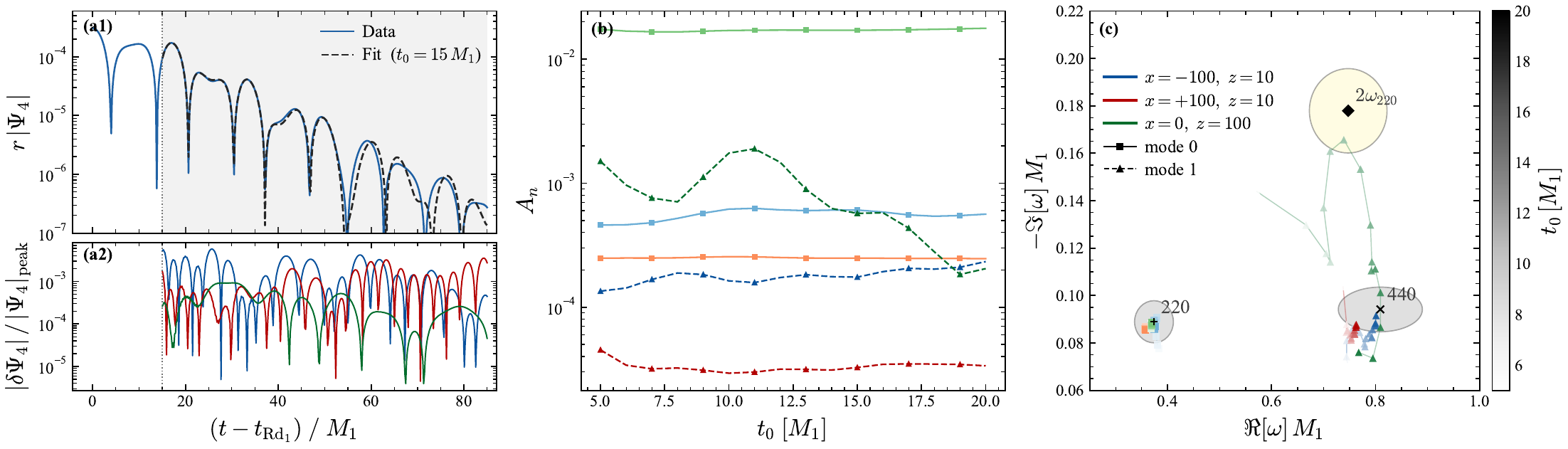}
    \caption{Direct ringdown following the first merger, from ID AU. \textbf{a1}: Signal (blue) and best-fit model (black) using $N=2$ free damped sinusoids. \textbf{a2:} Residuals between the signal and the best fit with the same model as in the panel above, at three different extraction points (see legend). \textbf{b:} Amplitude of the modes obtained fitting the signal as a a superposition of two free damped sinuosids, as a function of the start time of the fit, $t_0$. The recovered amplitudes of the dominant mode (lighter colors, solid lines) are very stable. The subdominant mode (darker color, dashed lines) is relatively stable as well, except for the extraction point $x=0,z=100$. \textbf{c:} Frequencies recovered from the free-frequency fits at different ringdown start times. There is clear evidence of two modes, compatible with the fundamental modes of $\ell=2$ and $\ell=4$, although this higher harmonic is recovered less accurately.  Similar results for other different initial data configurations are shown in Appendix~\ref{app:fitting_results}. 
    }
    \label{fig:mode_content}
\end{figure*}
We now will investigate in more detail the ringdown signal after the first merger. Distorted, isolated BHs are known to relax to an equilibrium state via the emision of ``ringdown'' waves, well approximated by a superposition of damped sinusoids~\cite{Berti:2009kk,Berti:2025hly}
\begin{equation}\label{eq:ringdown_model}
    \psi_4 \approx \frac{1}{r}\sum_{\ell m n} A_{\ell m n} e^{-i\omega_{\ell m n} t} \, , 
\end{equation}
where $A_{\ell m n},\omega_{\ell m n}$ are complex amplitudes and frequencies, depending on the angular mode $\ell$ and tone number $n$ (due to symmetry the azimuthal number $m$ can be set to zero). Note that theoretical calculations of the linearized modes of an isolated, ringing BH of mass $M$ yield for the fundamental $\ell=2$ and $\ell=4$~\cite{Berti:2009kk,Berti:2025hly} (equatorial symmetry imposes that the $\ell=3$ mode vanishes)
\beq
M\omega_{20}&=&0.373672 - 0.088962 i\,, \label{eq:omega_20}\\
M\omega_{40}&=&0.809178 - 0.094164 i \label{eq:omega_40} \,,
\eeq
respectively.

We expect the characteristic frequencies $\omega$ excited by the first merger M1 to be close, but possibly not exactly equal to, the ones above with $M=M_1$, since the merger is no longer isolated. In order to extract these frequencies, we find the best fit parameters in Eq.~\eqref{eq:ringdown_model} using \texttt{Jaxqualin}~\cite{Cheung:2023vki}. 
%\sout{We find the best fits are achieved for $N=2$, where the signal is compatible with the superposition of the fundamental $\ell=2,4$ modes -- we do not find significant evidence for overtone excitation \js{or quadratic modes}. }
We are able to extract $N=2$ modes accurately, demonstrating that the recovered frequencies and amplitudes are sufficiently stable. These two modes are typically consistent with the $\ell=2,4$ fundamental modes -- we do not find significant evidence for overtone or quadratic mode excitation.

A representative example of the fitted signal, mismatches, and the recovered amplitudes and frequencies is shown in Fig.~\ref{fig:mode_content}. We focus on ID AU, but the results are qualitatively similar for all the configurations considered here, see Appendix~\ref{app:fitting_results}. Notice also that we can recover quite accurately the amplitude of the dominant mode (the quadrupole), as shown in the middle panel of Fig.~\ref{fig:mode_content}. Moreover, when the signal is lensed (blue points in the figure), the amplitude of the dominant and subdominant modes are comparable. This is a consequence of frequency-dependent amplification, which we further explore in Sec.~\ref{sec:amp_from_lensing}.

%%%%%%%%%%%%%%%%%%%%%%%%%%%%%%%%%%%%%%%%%%%%%%%
\subsubsection{Doppler and gravitational shifts}
%%%%%%%%%%%%%%%%%%%%%%%%%%%%%%%%%%%%%%%%%%%%%%%%
%\jr{i moved the table to the appendix, since i think all the info of the table is basically already in the figure}

We expect the QNM frequencies observed to be gravitationally and Doppler shifted. In order to test if this is an observable effect in our simulations, we extract the frequency and damping time of the two dominant QNMs, at different observing points, for the four triplet configurations of Table~\ref{tab:nr_quantities}. We average the extracted frequencies for varying ringdown start times between $t_0-t_{\rm peak}\in[10,20]M_1$, and list our results
in Table~\ref{tab:frequencies} in Appendix~\ref{app:fitting_results}. We use the (ringdown starting-time) averaged frequencies for the rest of the analysis.

\begin{figure}
    \centering
    \includegraphics[width=\linewidth]{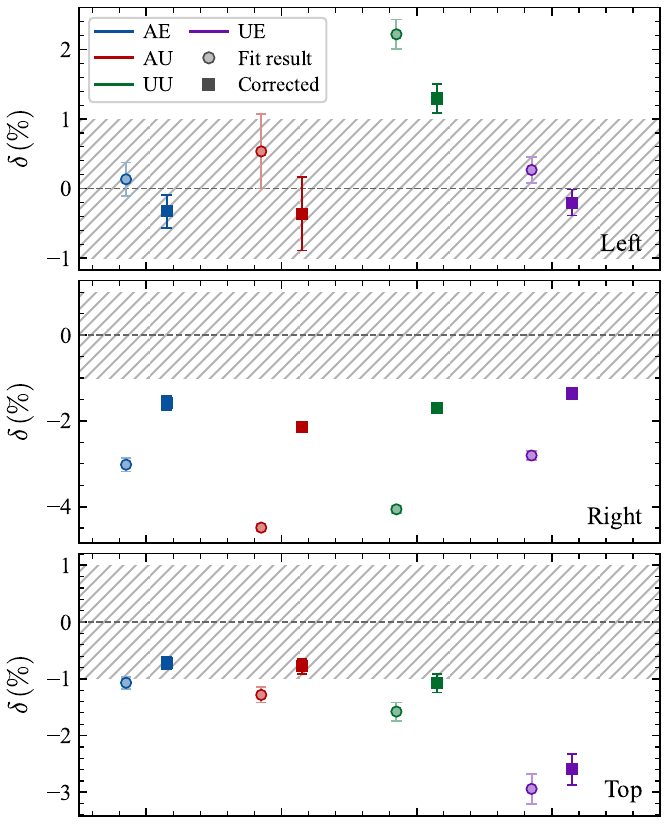}
    \caption{Fractional deviation from the theoretical value of the QNM frequency of isolated BHs, Eq.~\eqref{eq:delta_sigma}; We use a double damped sinusoid model for the first ringdown stage and the theoretical prediction for the dominant 20 mode frequency of a Schwarzschild BH, in ID AE, AU, UU, and UE.\\
    \noindent {\bf Top panel:} Observation point in the negative $x$-axis, $x=-100$, with $z=10$ (ID AE and AU), and $z=0$ (ID UU and UE); {\bf Middle panel:} Observation point $x=100$, with $z=10$ (ID AE and AU), and $z=0$ (ID UU and G);\\
    \noindent {\bf Bottom panel:} Observation point at $(x,z)=(0,100)$ for all runs. All distances are in code units. Shaded band shows $1\%$ deviation. We compare the frequencies directly extracted from the fit (light markers) and those corrected for gravitational and Doppler shifts according to Eq.~\eqref{eq:shift} (dark markers). We find reasonable agreement between the extracted frequencies and the theoretical prediction, with the correction always leading to better agreement.
    }
    \label{fig:deltas}
\end{figure}
Focus on the real part of the dominant extracted mode in the first ringdown phase -- this is the most accurately resolved frequency. We first compare this directly with the frequency of an isolated BH, Eq.~\eqref{eq:omega_20}. Then, we can correct the extracted value for the effects of Doppler and gravitational frequency shifts, using Eq.~\eqref{eq:shift} and Newtonian results for the relative position and motion of the BHs and observer. 

It is useful to define the fractional deviation of a given frequency from the isolated reference in Eq.~\eqref{eq:omega_20}
\begin{equation}
\delta = \frac{\rm Re [\omega ]- \rm Re [\omega _{\rm 20}]}{\rm Re [\omega _{\rm 20}]} \, .
\label{eq:delta_sigma}
\end{equation}
Figure~\ref{fig:deltas} shows this quantity for different observer positions, evaluating $\delta$ at $\omega$ as returned by the fit, or the corrected value~\eqref{eq:shift}. 
In general, we find that the extracted values agree with the shifted predictions significantly better, with all corrected values within 2.5\% of the theoretical prediction. It's also worth noting that observers to the right of the merger (positive $x$ direction) see {\it lower} frequencies, as a result of combined gravitational and Doppler \emph{red}-shifts. By contrast, observers to the left see gravitationally redshifted, but Doppler \emph{blue}-shifted signals, leading to a smaller deviation from the isolated BH frequency. In all cases, the deviations with respect to the predicted values of GR are in agreement with the red/blue-shift arising from the Doppler effect. Notice, also, that the Doppler shift estimates we make are an underestimation, as the remnant M1 suffers a short, but possibly non-negligible acceleration during the ringdown. 

%%%%%%%%%%%%%%%%%%%%%%%%%%%%%%%%%%%%%%%%%%%%%%%
\subsubsection{Amplification from lensing} \label{sec:amp_from_lensing}
%%%%%%%%%%%%%%%%%%%%%%%%%%%%%%%%%%%%%%%%%%%%%%%%
\begin{figure}
    \centering
    \includegraphics[width=\columnwidth]{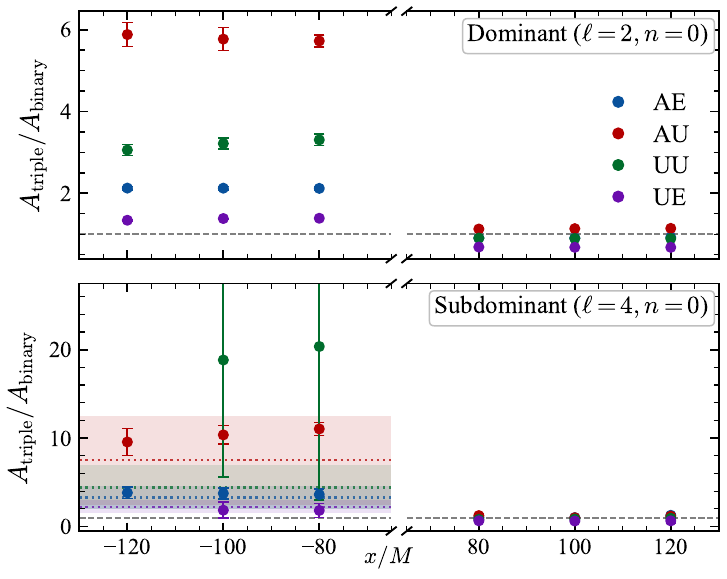}
    \caption{Ratio of amplitudes of the two free damped sinusoids fitted from the triple systems (ID AE, AU, UU, and UE) to that of reference binaries (ID AB and UB). The error bars are obtained by fitting at start times $(t_0-t_{\rm RD_1})/M_1=10,12.5,15,17.5,20$, and taking the mean and the standard deviation. The top panel (respectively, bottom panel) corresponds to the dominant (respectively, subdominant) mode. We omit the subdominant mode for ID UU and UE at $x=-120$, since its amplitude is very poorly constrained. Dashed bands are explained and discussed around Eq.~\eqref{eq:amplification_subdominant}.
    This result is clear evidence of magnification from the companion BH.
    }
    \label{fig:amplitude_ratio}
\end{figure}
Next, we compare the extracted QNM amplitudes of the first ringdown stage with those from the reference binaries -- ID AB and UB. These configurations are identical to initial data AE and AU (for ID AB), and UU and UE (for ID UB), except that there is no BH companion, $m_1=0$. Thus, one expects that the direct ringdown at extraction points with $x>0$ to be almost identical in amplitude, as the gravitational wave travels freely. On the other hand, the direct ringdown as observed at $x<0$ is expected to be different, due to gravitational lensing. 

Our results are summarized in Fig.~\ref{fig:amplitude_ratio}, and strongly support the interpretation above. For observers at $x>0$, the ratio between the amplitude measured from the triple system and from the reference binary is compatible with unity, for all initial data. When observing at $x<0$, we always find amplification. 

Upon more careful scrutiny, we observe two features: (i) data for which $m_1 \sim M_1$ (ID AU and UU) leads to a larger amplification relative to data for which $m_1 \sim M_1/2$ (ID AE and UE). Indeed, in the wave optics regime, the amplification $F$ of the signal scales as~\cite{1992grle.book.....S,Ezquiaga:2025gkd} 
\begin{equation}
    F^2\equiv \left(\frac{h^{\rm lensed}(\omega)}{ h^{\rm unlensed} (\omega)}\right)^2\propto m_1 \omega\sim \frac{m_1}{M_1} \, , \label{eq:amp_factor}
\end{equation}
where $h$ is the GW strain. Our analysis uses instead the Newman-Penrose scalar $\Psi_4 \sim \omega^2 h$, but since we focus on well defined frequency processes, this result holds identically for $\Psi_4$.
Hence, we expect ID AU, UU, to lead to a factor $\sqrt{2}$ greater amplification than ID AE, UE, respectively. 
Our results show a steeper dependence on the frequency, but the qualitative trend is compatible with that predicted from the wave optics limit of gravitational lensing. 
This is particularly apparent in Fig.~\ref{fig:amplitude_ratio} for the dominant modes;
(ii) the subdominant mode (see lower panel), which has a higher frequency, is significantly more amplified than the dominant mode. We may explain this on the basis of a dependence $F =\kappa m_1 \omega$, where $\kappa$ is independent of the frequency (it depends on the masses and the geometry). Then, we can predict the amplification of the subdominant mode from that of the dominant mode through 
\begin{equation}
    \frac{A^{\rm sub}_{\rm triple}}{A_{\rm binary}^{\rm sub}} \approx \frac{A^{\rm dom}_{\rm triple}}{A_{\rm binary}^{\rm dom}} \frac{\Re[\omega_{4,0}]}{\Re[\omega_{2,0}]} \, . \label{eq:amplification_subdominant}
\end{equation}
This behavior is shown as dashed bands in the bottom panel of Fig.~\ref{fig:amplitude_ratio}, showing very good agreement with our results. Notice that the error bars, especially for some instances of the subdominant mode, are relatively large, which limits how accurately we can infer the mode-by-mode amplification.

In summary, we find strong evidence of lensing in our results. Our results are self consistent, and qualitatively agree with perturbative results: higher frequencies get more amplified. Quantitatively, we deviate from the scaling of the amplification factor with frequency, since we find $F\propto m_1 \omega$. However, note that Eq.~\eqref{eq:amp_factor} is obtained in the standard weak gravity, thin lens, and small angle approximations, all of which certainly don't hold here. The triples we are studying probe a different, largely uncharted~\cite{Nambu:2019sqn,Willenborg:2023ixu} regime of lensing.

An improved model for lensing in this regime can be obtained building on the results of Ref.~\cite{Santos:2025ass}. This framework fixes the background to be that of the lens black hole ($m_1$ here), and calculates, in the linear regime, the gravitational waves produced by a monochromatic, time-varying mass quadrupole moment. The latter is held stationary at a finite distance from the lens, and is implemented using Dixon's formalism for extended mass distributions in curved spacetime~\cite{Dixon:1970zza,Dixon:1974xoz}. More details on this framework and its application to lensing will appear in~\cite{Santos:inprep}. This framework was developed to study sources that are orbiting binaries, but we straightforwardly implemented a linearly oscillating mass quadrupole moment to study the current setup. Using this model, we are able to recover the frequency scaling of the amplification in the ID AU and~AE runs, but not in ID UU and~UE. The latter are characterized by $m_1$ seeing the first ringdown face on, so the field near the lens in much stronger in these runs. It is an open question whether non-linear effects or the transient nature of ringdown play a role in this disagreement.

%
%%%%%%%%%%%%%%%%%%%%%%%%%%%%%%%%%%%%%%%%%%%%%%%
\subsubsection{Interference fringes} \label{sec:interference}
%%%%%%%%%%%%%%%%%%%%%%%%%%%%%%%%%%%%%%%%%%%%%%%%
Interference is one of the defining properties of wave mechanics. When the first ringdown scatters off the lens, we expect to see an interference pattern, resulting from different paths taken around $m_1$. This picture is clearly present in ID UU and UE. 

In Fig.~\ref{fig:fringes} we show a snapshot of the curvature scalar $\Psi_4$ for ID UU where interference fringes, resulting from destructive interference, are clearly visible (marked by black dashed lines). We find fringes both for the direct wave or first image (left) and the echo or second image (right) -- the fringes arise in this case from retrolensing. The fringes have the same aperture on both sides, with half angle $\theta_f \approx 33^\circ $. This is expected as in ID UU $m_1 = M_1$, so they behave as identical lenses. 
In contrast, for ID UE we measure the aperture angle of the fringes to be different on the left $\theta_f^{\rm left} \approx 48^\circ$, and on the right $\theta_f^{\rm right}\approx 33^\circ$, since the mass of the two BHs (the companion, and the first remnant) are now different, $m_1 < M_1$. 
Complete snapshot movies for both runs are available at \cite{YoutubeCanuda,web:CoG}.

In typical lensing analysis, the aperture of the fringes is expected to scale as $1/m_1 \omega$, ~\cite{Ezquiaga:2025gkd}. For ID UU and ID UE this would yield a fringe aperture ratio of $\sim 2$. However, as pointed out in the previous subsection, this setup is far from conventional. We can apply the more sophisticated lensing model~\cite{Santos:2025ass}, described in the end of Sec.~\ref{sec:amp_from_lensing}, to this problem.
Doing so, we recover the fringe aperture ratio of $\sim 1.45$ we observed in the numerical simulations. The model is incapable of predicting the individual apertures $\theta_f$.
\begin{figure}
    \centering
    \includegraphics[width=\linewidth]{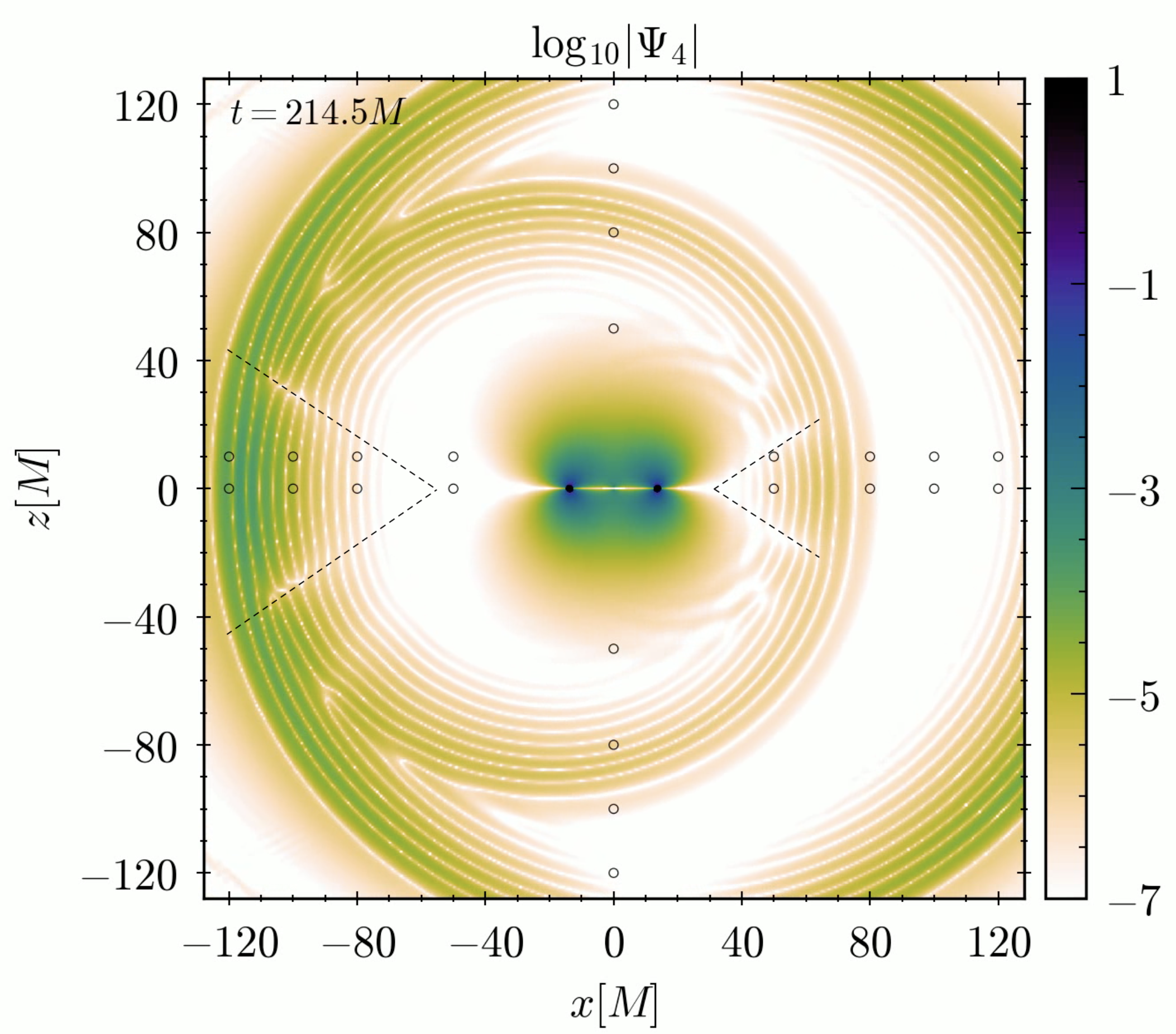}
    \caption{Snapshot of $|\Psi_4|$ in the run ID UU. The two black holes, $m_1$ and $M_1$ are marked by black dots and circles indicate different extraction points. The outer spherical front is the direct wave from the first merger. On the left, we see interference fringes (black dashed lines) after scattering off $m_1$. The scattering produces a new spherical wavefront (the echo). The echo also shows fringes on the left, at the same angle as the direct wave on the right, and resulting from scattering off $M_1$. A third spherical wavefront (second echo) is seen emerging from $M_1$. Videos of the numerical simulations can be found in~\cite{YoutubeCanuda, web:CoG}.
    }
    \label{fig:fringes}
\end{figure}
%
%%%%%%%%%%%%%%%%%%%%%%%%%%%%%%%%%%%%%%%%%%%%%%%
\subsubsection{Other modes}
%%%%%%%%%%%%%%%%%%%%%%%%%%%%%%%%%%%%%%%%%%%%%%%%
We have also investigated the presence of nonlinear modes in the signal. By considering a hybrid model, where we fix the signal to contain the $\ell=2,4$ fundamental modes, and one additional free frequency, we find some evidence of a quadratic QNM with frequency $\omega \sim 2\omega_{2,0}$. More accurate simulations and parameter estimation techniques will be necessary in order to fully unveil the role of nonlinear effects in mergers in triple systems. From our analysis we can only conclude that nonlinear QNMs, or resonantly excited harmonics, do not seem to play a key role. Notice that the radiation emitted in a head-on merger is small~\cite{Sperhake:2011ik,Witek:2010xi}. Quasi-circular inspirals, where a fraction of the radiation is emitted quasi-monochromatically for a longer time, may be better ``tuning forks'' to resonantly excite ringdown modes of the companion, or enhance nonlinear excitation of higher harmonics.

% 
%%%%%%%%%%%%%%%%%%%%%%%%%%%%%%%%%%%
\subsection{Second image\label{sec:lensing}}
%%%%%%%%%%%%%%%%%%%%%%%%%%%%%%%%%%%

As we noted, our results also show clear evidence for echoes from the first merger. In other words, observers see a direct ringdown from merger M1, but some of the radiation as it scatters and lenses close to the companion BH $m_1$, can deflect back to the observer. This produces a distorted copy of the first signal which can either be stronger or weaker depending on the observation point. In fact there should be an infinite number of such copies, as radiation travels more than once around the BH, but those signals are exponentially suppressed. Thus, we focus on the first echo, seen in the right panel of Fig.~\ref{fig:summary}.

%%%%%%%%%%%%%%%%%%%%%%%%%%%%%%%%%%%%%%%%%%%%%%%
\subsubsection{Time delay between images}
%%%%%%%%%%%%%%%%%%%%%%%%%%%%%%%%%%%%%%%%%%%%%%%%

The observed delay time between main direct pulse and second image can be compared against a simple prediction in the geometric optics. Take a gravitational wave as a particle on a null geodesic, with quasinormal modes corresponding to trapped particles in the light ring~\cite{Cardoso:2008bp,Berti:2009kk}. For an observer on the $z$ axis, the time delay is the sum of two components:
\textit{(i)} the travel time from the first merger remnant $M_1$ to the companion $m_1$, $\Delta t_{M_1\to m_1}$;  \textit{(ii)} the time for the null particle to go half a loop around the light ring of $m_1$, $\Delta t_{m_1}$.

%
%\begin{figure}
%    \centering
%\includegraphics[width=1.0\linewidth]{figures/GO_sketch.png}
%    \caption{A cartoon of the geometric optics approximation for the time delay of the echo seen by an observer on the $z$ axis. The echo has a time delay given roughly by the travel time from one light ring to that of the companion and then traveling half a light ring orbit. \js{I think we can remove this...}}
%    \label{fig:GO_sketch}
%\end{figure}
%
The first may be approximated by considering that both $M_1$ and $m_1$ are static isolated Schwarzschild BHs, whose spacetimes are glued at their Newtonian equipotential point. We then time a radial null geodesic from the light ring of $M1$ to the equipotential point and then down to the light ring of $m_1$. If the BHs are at a distance $d$, this can be shown (proof in Appendix~\ref{app:time_delay}) to equal to
\begin{equation}
    \Delta t_{M_1\to m_1} \approx 2 \int_{3M} ^{d} dr \left(1-{2M}/{r}\right)^{-1} \, , \label{eq:time_delay_1}
\end{equation}
where $M=m_1+M_1$ is the total mass. The second component is just half a period for a light ring orbit
\begin{equation}
    \Delta t_{m_1} \approx  \pi \ 3 \sqrt 3 \ m_1 \, .
\end{equation}
When an observer is to the right of the merger (meaning large positive $x$), the time delay is obtained identically, but with the null particle taking a full orbit around $m_1$ and then traveling back to $M_1$ and traveling a half orbit around its light ring. The distance $d$ between the BHs are estimated from Newtonian calculations. 
\begin{table}[]
\begin{tabular}{lcccc}
\hline 
\hline 
ID & Observer    &$\Delta t_{\rm num}/M$ & $\Delta t_{\rm go} / M$  & $t_d/M$ \\
\hline 
UU  & $(0,0,100)$ & $52 \pm 8$            &  52  & $49.0$  \\
   & $(100,0,0)$ & $95\pm 8$             &  107 & $90.8$ \\
UE  & $(0,0,100)$ & $37 \pm 11$           &  50  & $33.2$  \\
   & $(100,0,0)$ & $92 \pm 11$           & 103  & $87.8$ \\
\hline 
\hline
\end{tabular}
\caption{Time delay between the peak in the direct ringdown following the first merger $t_{\rm RD_1}$, and the secondary ringdown or echo $t_{\rm RD_2}$, defined as $\Delta t_{\rm num} = t_{\rm RD_2}-t_{\rm RD_1}$, in code units, for two different initial configurations and observing points. Errors are given by a period of the wave, due to uncertainty in defining the peak of the waveform. The middle column provides the theoretical prediction from the geometric optics limit, $\Delta t_{\rm go}$. The last column indicates the time delay obtain by minimizing the mismatch between the first and second images, see~\eqref{eq:images} and Fig.~\ref{fig:delay}.}
\label{tab:delay}
\end{table}
Our Newtonian and geometric optics estimates, together with the time delay as estimated from the peak of the direct and echo waveform, are shown in Table~\ref{tab:delay}.

Despite its simplicity, our model is always within $\sim 10\%$ of the interval for the observed time delay. The geometric optics calculation always overestimates $\Delta t$ since it also overestimates the length of the path traveled by the null particle. Recall that the geometric optics limit is a good approximation in the high frequency regime, $m_1 \omega \gg1 $, which explains the better agreement for ID UU. Overall, since we have $m_1 \omega \lesssim 1 $, the agreement found in Tab.~\ref{tab:delay} is remarkable. 

%%%%%%%%%%%%%%%%%%%%%%%%%%%%%%%%%%%%%%%%%%%%%%%
\subsubsection{Properties of the lensed ringdown}
%%%%%%%%%%%%%%%%%%%%%%%%%%%%%%%%%%%%%%%%%%%%%%%%
\begin{figure*}
    \centering
    \includegraphics[width=\linewidth]{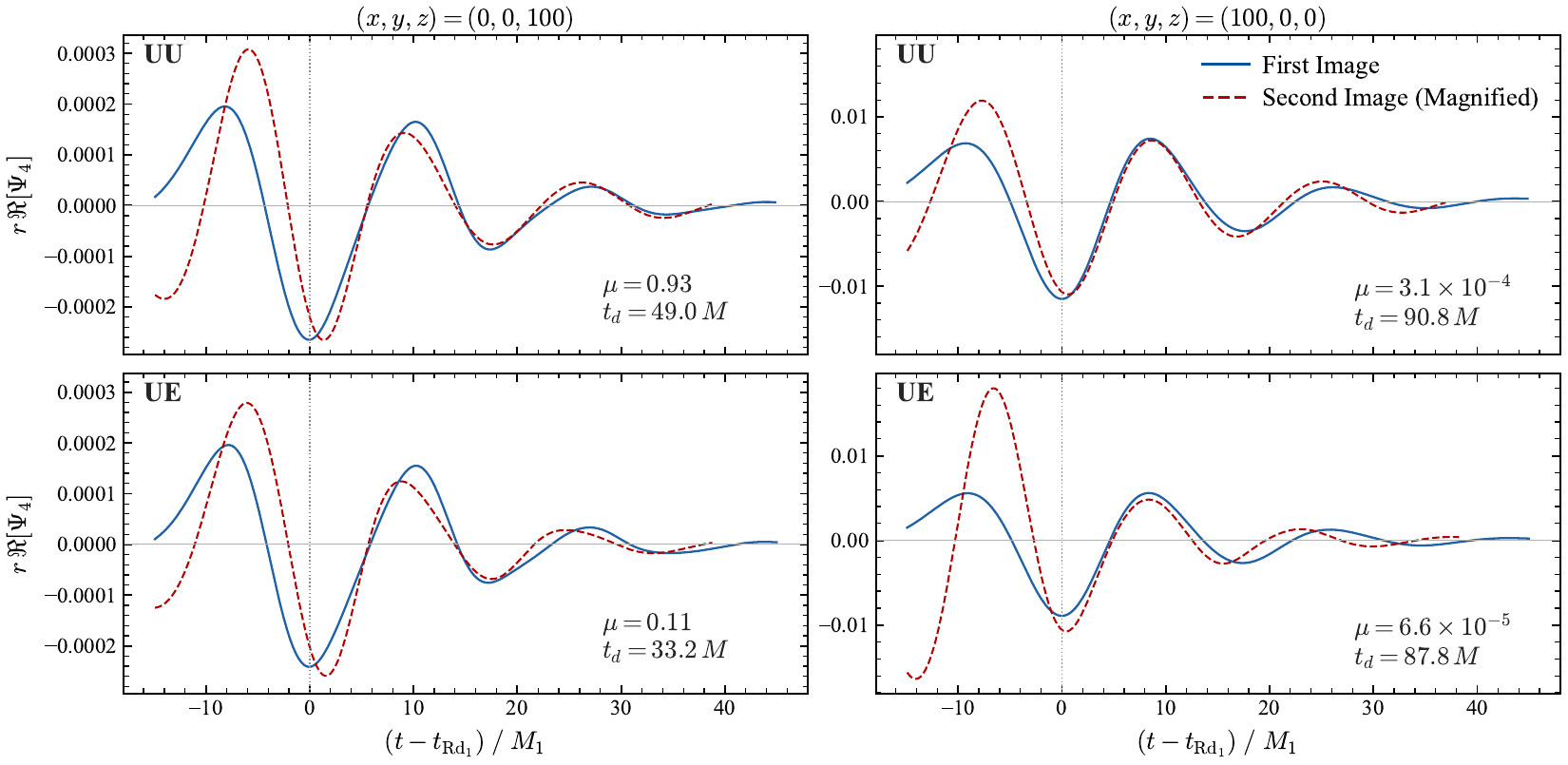}
    \caption{First ringdown (blue), and lensed ringdown (dashed, red), after finding the best-fit magnification $\mu$ and time delay $t_d$ to align both signals. The upper (respectively lower) panels correspond to ID UU (respectively, UE). The left and right columns correspond to the indicated observing points: $(x,y,z) = (0,0,100)$, and $(100,0,0)$, respectively. }
    \label{fig:delay}
\end{figure*}
We examine further the secondary image in Fig.~\ref{fig:delay}. The smaller amplitude of this signal hinders an analysis of the amplification akin to Sec.~\ref{sec:amp_from_lensing}. Thus we instead employ a simpler analysis based entirely on the geometric optics limit. Notice that the ringdown is quite monochromatic --- in the frequency domain, all frequencies with $\omega > \omega_{2,0}$ are suppressed exponentially by the BH reflectivity~\cite{Berti:2025hly}. In the geometric optics limit, when multiple images are present, we expect the secondary image to be simply a (de)magnified and delayed copy of the first ringdown. If we denote as $h_1(t) = r \Re [\Psi_4]$ the signal of the first ringdown, and $h_2$ the same quantity, but for the second image, we propose a relation of the form
\begin{equation}\label{eq:images}
    h_2 \approx \sqrt{\mu} h_1(t+t_d) \, ,
\end{equation}
where $\mu$ is the physical magnification (i.e., the ratio of the flux carried by each image)~\footnote{Notice that the magnification is related to the amplification factor in the geometric optics regime as $\mu=|F^2|$. Note also that the nearly monochromatic nature of the ringdown allows us to model the echo as having a single magnification (rather than a sum of differently magnified frequency components).}, and $t_d$ the time delay. We minimize the mismatch between both signals to find the best fit values for the magnification $\mu$ and the time delay $t_d$, and report their values in Fig.~\ref{fig:delay}. 

First, notice that the time delays are very similar to the peak-amplitude delays as reported in Table~\ref{tab:delay}. Second, the agreement between the unlensed (blue), and the lensed waveforms (red), is quite good, despite the simplicity of the model. Indeed, we are beyond the regime where Eq.~\eqref{eq:images} is expected to be applicable: the lens is not thin, the waves impinging the lens are not plane-fronted, the frequency of the ringdown waves is comparable to the mass of the lens, and strong, nonlinear field effects could be present. Nevertheless, it leads to a good alignment between the two ringdown images: in all cases we achieve a mismatch between the direct ringdown and the lensed signal of $\sim 10^{-2}$. 

Moreover, the magnification $\mu$ contains some interesting information. When observing along the positive $x$-axis, the second image is very demagnified, $\mu \sim 10^{-4}$. This is expected -- the signal has to loop once around the BH, and hence its amplitude is suppressed. Indeed, signals are expected to be exponentially suppressed with the number of loops around the BH that they perform~\cite{Johnson:2019ljv}.
%we are precisely showing the analogue of looking straight at a BH and seeing the back of your own head. This corresponds to a $n=2$ photon-ring, and we know that the intensity of each successive photon ring is demagnified exponentially. 
A crude estimate for the magnification can be done using the geometry of the problem. If the radiation were high frequency, the fraction of rays being deflected back by $m_1$ would scale like the region occupied by $m_1$ in the sky~\cite{Cardoso:2019dte}, i.e., $\mu \sim m_1^2/d_0^2$, with $d_0$ the initial distance between $m_2$ and $m_1$. If we also account for the different emission at different angles, as estimated by the leading $\ell=2,m=0$ spin-weight $s=-2$ spherical harmonics, this simple estimate reproduces both the order of magnitude of our results and the observed dependence on $m_1$ for UU and UE initial data.

The polarization content of the first and second images might not be the same due to helicity dependent scattering, a well known signature of gravitational lensing~\cite{ Pijnenburg:2024btj,Chan:2025wgz,Oancea:2022szu}. More interestingly, if the lens is rotating, the polarization mixing can be significantly enhanced~\cite{Kubota:2024zkv, Li:2022izh, Saketh:2025cwf}. This analysis is outside the scope of this work.

%On the other hand, the magnification is much larger when looking along the $z$-axis, when the signal does only half a loop around the BH. Finally, we also have that generically \vc{were these defined?} $\mu_{G} \ll \mu_F$. One possible explanation is that ID UU is resonant, since the mass of the lens and the mass of the ringing BH are comparable, $m_1 \approx M_1$, so the secondary image may be further amplified. However, the ringdown of non-spinning BHs is likely too short-lived to resonantly excite the companion; a more likely explanation is simply provided by Eq.~\eqref{eq:amp_factor}, the companion is simply more massive.

%%%%%%%%%%%%%%%%%%%%%%%%%%%%%%%%%%%%%%%%%%%%%%%
\subsubsection{New modes in secondary images}
%%%%%%%%%%%%%%%%%%%%%%%%%%%%%%%%%%%%%%%%%%%%%%%%
We can now examine the second ringdown in more detail. As evidenced by Fig.~\ref{fig:delay}, this second image is not an exact copy of the first image -- we expect this to be manifest in its mode content. Theoretically, we expect two things: (i) the first image should have modes that are redshifted, when extracting in the positive $x$-axis, whereas the second image modes should be blueshifted instead, and (ii) if resonant excitation is taking place, albeit by a small amount, we should observe a superposition of the (blueshifted) fundamental mode of the remnant $M_1$, and the fundamental mode of the lens $m_1$. Our results confirm that this is, indeed, the case. 

We proceed as in the previous section and model the second ringdown with $N=2$ free damped sinusoids. We notice that this is somewhat more challenging, due to uncertainties in the numerical simulations stemming from back-reflection from the refinement boundaries. In particular, we choose to be conservative, and model only the first $t-t_{\rm Rd_2} < 65M_1$, which we are confident is unaffected by numerical noise. The results from the fit, in the manner of Fig.~\ref{fig:mode_content}, are shown in Fig.~\ref{fig:second_fits} in Appendix~\ref{app:fitting_results}. We summarize the key results in Table~\ref{tab:second_ringdown}. In order to obtain those values, we average over the start time of the ringdown in the range $t_0 \in [6,20]M_1$, to estimate the uncertainty. 

%\begin{table}[t]
%\begin{ruledtabular}
%\begin{tabular}{llcccc}
%  ID & Observer
%     & $\Re[\omega_0]M_1$ & $-\Im[\omega_0]M_1$
%     & $\Re[\omega_1]M_1$ & $-\Im[\omega_1]M_1$ \\
%\hline
%  \multirow{3}{*}{UU} & Top & $0.378(1)$ & $0.074(2)$ & $0.6(1)$ & $0.09(6)$ \\
%   & Right & $0.35(3)$ & $0.09(1)$ & $0.41(1)$ & $0.064(9)$ \\
%\hline
%  \multirow{3}{*}{UE} & Top & $0.395(1)$ & $0.088(2)$ & $0.799(6)$ & $0.096(2)$ \\
%   & Right & $0.432(3)$ & $0.091(1)$ & $0.8(1)$ & $0.09(1)$ 
%\end{tabular}
%\end{ruledtabular}
%\caption{Frequencies inferred from $N=2$ free-mode fits of the second ringdown as observed from the top ($x=y=0,z=100$), and from the right ($x=100,y=z=0$) of the lens. We average the results obtained when varying the ringdown start time $(t_0 - t_{\rm Rd_1})/M_1 \in [6,16]M_1$. The value in between brackets indicates the uncertainty in the last digit. \vc{I don't like this way of presenting errors, and i think the 0.8 entry is wrong?}
%  \label{tab:second_ringdown}
%}
%\end{table}
%
\begin{table}[t]
\begin{ruledtabular}
\begin{tabular}{llcc}
  ID & Observer & $M_1\omega_0$       & $M_1\omega_1$ \\
\hline
  \multirow{3}{*}{UU} 
    & Top      & $0.378(1)-i0.074(2)$ & $0.6(1)-i 0.09(6)$ \\
   & Right     & $0.35(3)-i0.09(1)$   & $0.41(1)-i0.064(9)$ \\
\hline
  \multirow{3}{*}{UE} 
  & Top       & $0.395(1)-i0.088(2)$ & $0.799(6)-i0.096(2)$ \\
& Right       & $0.432(3)-i0.091(1)$ & $0.8(1)-i0.09(1)$ 
\end{tabular}
\end{ruledtabular}
\caption{Frequencies inferred from $N=2$ free-mode fits of the second ringdown as observed from the top ($x=y=0,z=100$), and from the right ($x=100,y=z=0$) of the lens. We average the results obtained when varying the ringdown start time $(t_0 - t_{\rm Rd_1})/M_1 \in [6,20]M_1$. The value in between brackets indicates the uncertainty in the last digit. 
  \label{tab:second_ringdown}
}
\end{table}
Look first at ID UE. In this case, $M_1 \neq m_1$, so we do not expect resonant excitation of modes to play a significant role, especially given how short the ringdown is. The first mode extracted corresponds to a blueshifted fundamental $\ell=2,n=0$ mode. The frequency shift is most evident when extracting at the right, as this echo is blueshifted twice: at emission and at scattering. Using the results of Sec.~\ref{subsec:Doppler} and our Newtonian calculations, we estimate $\omega_{\rm obs}/\omega_{\rm em} \approx 1.06$. This is not enough to reach the $\mathcal{O}(1\%)$ agreement obtained in  Table~\ref{eq:delta_sigma} for the first ringdown. This is partly explained by the echo carrying more imprints of strong field and relativistic effects, not captured by our Newtonian estimates.
%\jr{In this case, the image is also ``retrolensed'' -- as it propagates back from the lens to the observer, it crosses the BH...} 
The second harmonic is inferred less accurately, but is in good agreement with the $\ell=4,n=0$ mode, after accounting for blueshift effects. 

The situation for ID UU is quite different. Now, $M_1 \approx m_1$. Even though the ringdown is very short -- and hence, not sufficient to drive resonant excitation of QNMs to a large amplitude -- it may be strong enough to subtly excite some of the modes of the companion BH. Indeed, this seems to be what we observe, especially when analyzing the signal to the \emph{right} of the observer: we measure two modes, both close to the frequency of the $\ell=2,n=0$ mode, which justifies the high uncertainties in the extracted values. We interpret the first mode, $\omega_0$, to be a mode of the lens $m_1$, which gets blueshifted by the motion of the lens, so $\omega_{\rm obs}/\omega_{\rm em} \approx 1.05$. The second, $\omega_1$, would be the scattered $\ell=2$ mode of the first merger, which gets blueshifted twice, as in ID UE, so $\omega_{\rm obs}/\omega_{\rm em} \approx 1.07$. The extracted values are compatible with this interpretation within their uncertainties. Moreover, this explanation is motivated by the high blueshift observed for the first mode in ID UE, completely incompatible with the first mode in ID UU. 
%\jr{actually, slightly redshifted, $m_1$ is moving towards $M_1$} \js{this should also be blueshifted}
When extracting at the top, we don't see the lens mode as the observer is almost exactly edge on. Instead, we extract two modes compatible with scattered $\ell=2,4$ $n=0$ modes, as in ID UE.

The analysis of the echoes shows that their time delays, amplitudes and frequencies are compatible with analytical approximations and with the previous results for the direct ringdown in Sec.~\ref{sec:direct}. Most interestingly, they show tentative evidence of resonant excitation of modes of the lens $m_1$, both at the level of the amplifications (Fig.~\ref{fig:delay}) and of the frequencies (Tab.~\ref{tab:second_ringdown}). However, the echo is weak, its analysis is complex and there are no perturbative models suited to this setup we can benchmark against. Thus, resonant effects are merely a tentative explanation.
%\jr{say some words of caution to express that this analysis is complex -- these are our results and the interpretation we give, but it does not need to be the final word.}

%%%%%%%%%%%%%%%%%%%%%%%%%%%%%%%%%%%%%%%%%%%%%%%
\section{Discussion} \label{sec:discussion}
%%%%%%%%%%%%%%%%%%%%%%%%%%%%%%%%%%%%%%%%%%%%%%%

Third body companions are reasonable astrophysical environments, they will affect the GW signal in ways that in principle can be modeled analytically. However, in the absence of a clear hierarchy of distances or masses in the triple system, nonlinear effects, or effects that go beyond the usual lensing assumptions, are likely to be present. Numerical simulations such as those carried here will be informative to model such events.

We studied the GW signal from merging triple systems. Our methods are straightforward applications of numerical relativity tools, but uncover a number of interesting aspects and lessons. The first take-away is that GWs {\it are} lensed, magnified, Doppler and gravitational shifted in ways that we control and understand. Close-by companions affect propagation of GWs.

The second lesson is that tides are always present but their effects on light ring properties and ringdown are orders of magnitude suppressed as compared to other leading-order effects. Companions can induce Doppler shifted secondary images or resonant frequencies, which shows up as ringdown with double close-by frequencies (and which might be wrongly interpreted as isospectrality break-up). Clear evidence for resonant excitation possibly requires spinning BHs, for which ringdown is longer lived, or a binary in circular motion; this would be an interesting follow up to this work. 

Our results answer the questions posed in the Introduction: (i) the ringdown is modified in the presence of a companion; (ii) we don't find significant focusing. We tested other initial data with closer BHs, to test for violent processes and BH formation by lensing of radiation~\cite{Blas:2024ypz}. We find no evidence for such phenomena; (iii) we do not find evidence of enhanced nonlinearities. Our mode analysis of the ringdown is well explained by a combination of linear quasinormal modes of each BH. Other geometries, or mass ratios, may be more efficient at enhancing nonlinear dynamics. We also searched for low-frequency modes, that could be binary BH modes~\cite{Assumpcao:2018bka,Bernard:2019nkv}, but found no convincing evidence for their presence.

%%%%%%%%%%%%%%%%%%%%%%%%
\begin{acknowledgments}
%%%%%%%%%%%%%%%%%%%%%%%%
%
The Center of Gravity is a Center of Excellence funded by the Danish National Research Foundation under grant No. DNRF184.
We acknowledge support by VILLUM Foundation (grant no. VIL37766).
V.C.\ is a Villum Investigator.  
V.C. and J.S.S acknowledge financial support provided under the European Union’s H2020 ERC Advanced Grant “Black holes: gravitational engines of discovery” grant agreement no. Gravitas–101052587. 
Views and opinions expressed are however those of the author only and do not necessarily reflect those of the European Union or the European Research Council. Neither the European Union nor the granting authority can be held responsible for them.
This project has received funding from the European Union's Horizon 2020 research and innovation programme under the Marie Sklodowska-Curie grant agreement No 101007855 and No 101131233.
The authors thank the Fundação para a Ciência e Tecnologia (FCT), Portugal, for the financial support to the Center for Astrophysics and Gravitation (CENTRA/IST/ULisboa) through grant No. UID/PRR/00099/2025 and grant No. UID/00099/2025.
This work is supported by Simons Foundation International \cite{sfi} and the Simons Foundation \cite{sf} through Simons Foundation grant SFI-MPS-BH-00012593-11.
G.F. gratefully acknowledges the support of University of Calabria through a research fellowship funded by DR 1688/2023.
The Tycho supercomputer hosted at the SCIENCE HPC center at the University of Copenhagen was used for supporting this work.
Authors acknowledge computational resources for simulations and analysis from the Alarico HPPC Computing Facility at the University of Calabria.
J.~R.-Y. is supported by NSF Grants No.~AST-2307146, No.~PHY-2513337, No.~PHY-090003, and No.~PHY-20043, by NASA Grant No.~21-ATP21-0010, by John Templeton Foundation Grant No.~62840, by the Simons Foundation [MPS-SIP-00001698, E.B.], by the Simons Foundation International [SFI-MPS-BH-00012593-02], and by Italian Ministry of Foreign Affairs and International Cooperation Grant No.~PGR01167.

\end{acknowledgments}

%\clearpage
\clearpage 
\appendix
\onecolumngrid

%%%%%%%%%%%%%%%%%%%%%%%%%%%%%%%%%%%%%%%%%%%
\section{Convergence of numerical simulations}\label{app:convergence}
%%%%%%%%%%%%%%%%%%%%%%%%%%%%%%%%%%%%%%%%%%%
%
\begin{figure*}[ht!]
    \centering 
    \includegraphics[width=.49\linewidth]{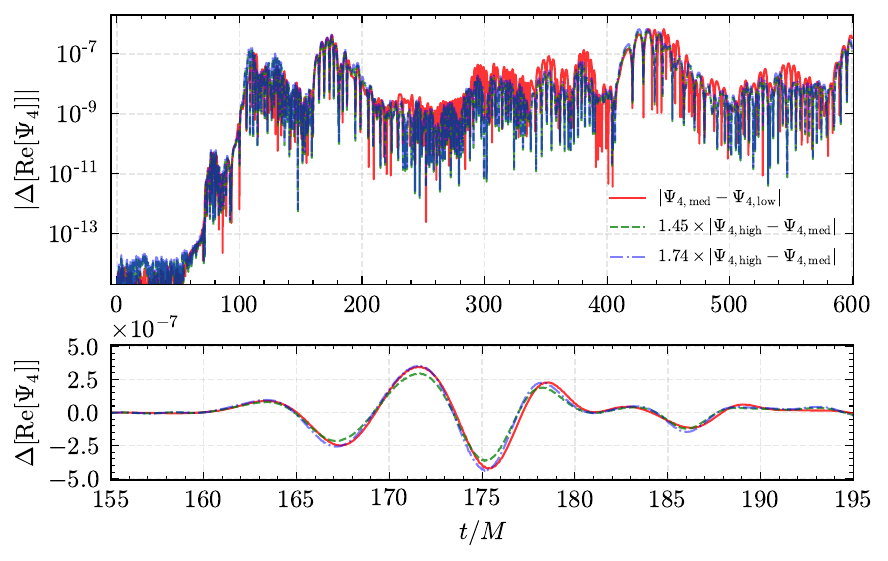}
    \includegraphics[width=.49\linewidth]{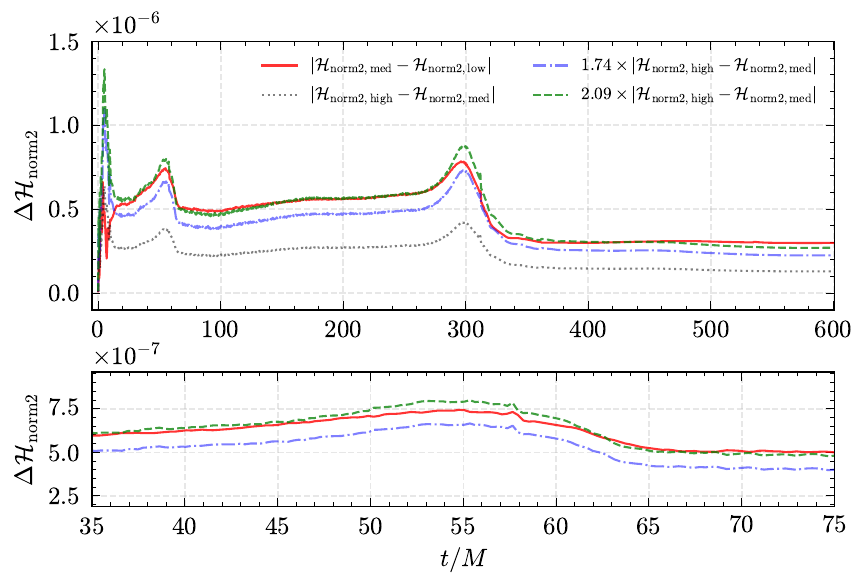}
    \caption{Convergence test for run ID AE: in Table~\ref{tab:nr_quantities}. Left column displays $\Psi_4$ at the observer location $(x,y,z)=(0,0,100)M$, while the right column the L-2 norm of the Hamiltonian constraint. Top panels show the entire evolution and bottom panels provide a zoomed-in window around the first merger. Convergence factors for second, third and fourth order convergence are 1.45, 1.74 and 2.09.}
    \label{fig:convergence}
\end{figure*}
To ensure that our simulations are performed at a sufficiently high resolution to enter the convergence regime, we perform convergence tests. To this end, we run the configuration ID AE in Table~\ref{tab:nr_quantities} at three different spatial resolutions, which we denote here as $\Delta x_{\rm{high}} < \Delta x_{\rm{med}} < \Delta x_{\rm{low}}$. For a given quantity $h$, we compute the differences $\Delta$ between the fine and medium resolution results, $|h_{\rm{high}} - h_{\rm{med}}|$, as well as between the medium and coarse resolution results, $|h_{\rm{med}} - h_{\rm{low}}|$. If the solution converges at $n$-th order, these residuals scale with the numerical error according to the relation discussed in \cite{alcubierre2008introduction}
\begin{align}
\label{eq:convergence_factor}
\frac{|h_{\rm{med}} - h_{\rm{low}}|}{|h_{\rm{high}} - h_{\rm{med}}|}
& = Q_n \equiv
\frac{\Delta x_{\rm{med}}^n - \Delta x_{\rm{low}}^n}{\Delta x_{\rm{high}}^n - \Delta x_{\rm{med}}^n}
\,,
\end{align}
where $Q_n$ is the $n$-th order convergence factor for the resolutions considered. We run setup ID AE: in Table~\ref{tab:nr_quantities} at three different resolutions $\Delta x_{\rm{high}} \simeq 0.58 M$, $\Delta x_{\rm{med}} \simeq 0.69 M$, $\Delta x_{\rm{low}} = 0.83M$ defined on the extraction refinement level, and examine if the criterion in Eq.~\eqref{eq:convergence_factor} is satisfied. The corresponding convergence factors are $Q_2 = 1.45, Q_3 = 1.74$ and $Q_4 = 2.09$ for second, third and fourth order convergence, respectively. 

All simulations presented in this work adopt fourth-order finite difference stencils to compute spatial derivatives, combined with a fourth-order Runge-Kutta scheme for time integration. The interpolation across refinement level boundaries, however, is performed using a second-order accurate stencil in time and a fifth-order accurate one in space. Given this combination of numerical methods, we anticipate a mixed convergence order.

The convergence test for the real part of Newman-Penrose scalar $\Psi_4$ extracted at $(x,y,z)=(0,0,100)M$ is shown in the left panel of Fig.~\ref{fig:convergence}. Overall, the waveforms exhibit over-convergence across most of the evolution. Around the merger time, the convergence order drops to third order, as shown in the bottom panel of the same plot, which shows a zoom-in around the first merger. This behavior is not unexpected, given that the head-on collisions considered here generate a strong signal only when the BHs closely approach merger. This stands in contrast to long inspirals, where the gravitational wave signal persists over many cycles and the convergence can be studied over a longer time scale.

To obtain a more reliable estimate of the convergence order, we also examine the L-$2$ norm of the Hamiltonian constraint $\mathcal{H_{\rm{norm2}}}$. Convergence test for the latter is displayed in the right panel of Fig.~\ref{fig:convergence}. We observe fourth-order convergence across most of the evolution, decreasing to between third and fourth order near the merger times, which correspond to growing peaks in the evolution over time. Additionally, we provide a zoom-in on the first merger, where this feature becomes manifest.

Given the results presented here, we choose to perform all the simulations with the medium resolution on the extraction level $\Delta x_{\rm{med}} \simeq 0.69 M \equiv \Delta x_{\rm{ext}}$.

%%%%%%%%%%%%%%%%%%%%%%%%%%%%%%%%%%%%%%%%%%%
\section{Additional Fitting Results}\label{app:fitting_results}
%%%%%%%%%%%%%%%%%%%%%%%%%%%%%%%%%%%%%%%%%%%
%
\begin{figure*}[h!]
    \centering
    \includegraphics[width=\linewidth]{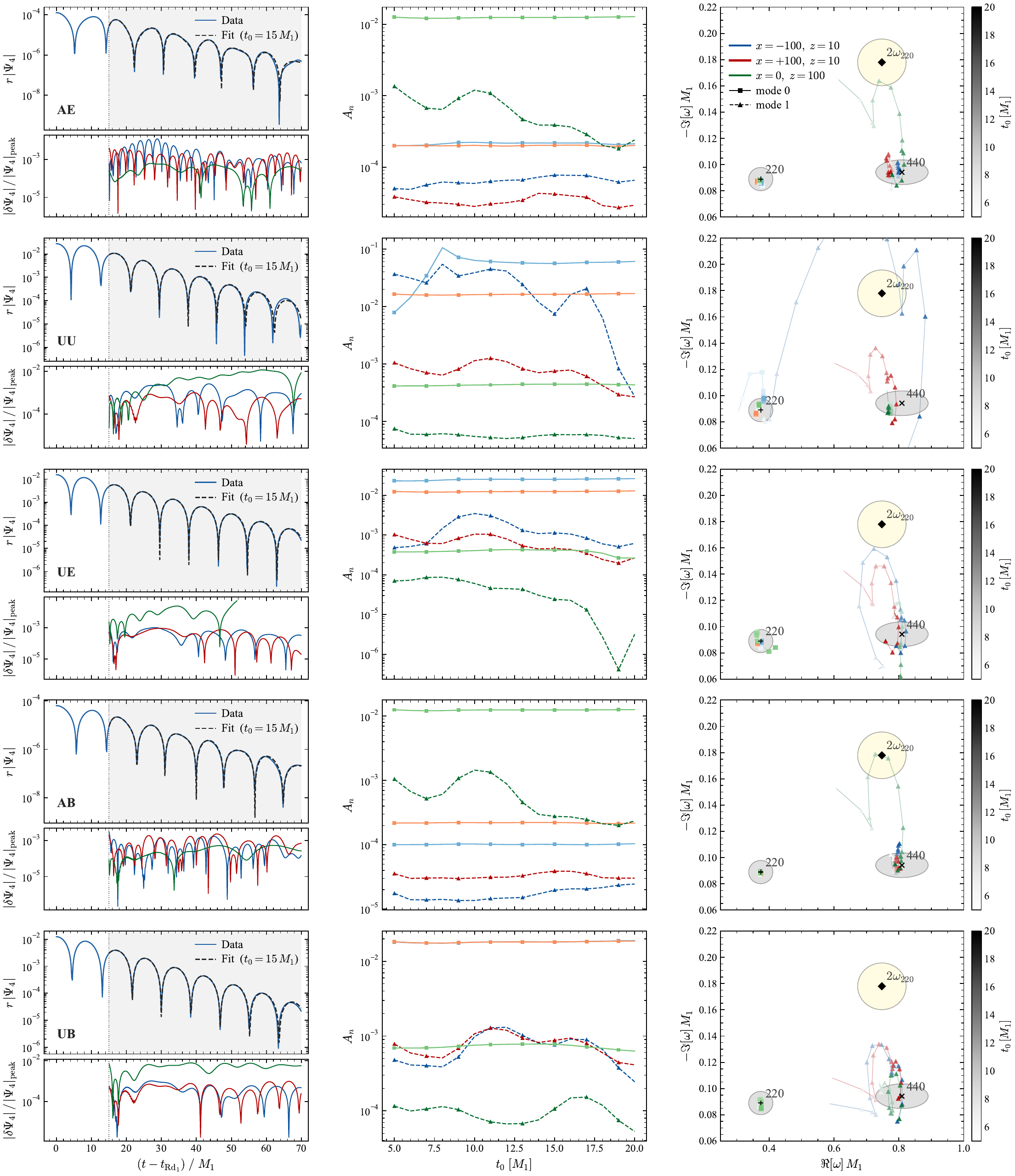}
    \caption{Same content as in Fig.~\ref{fig:mode_content}, but for initial data AE, UU, UE, AB and UB. 
    }
    \label{fig:modes_extra}
\end{figure*}
Here we summarize the results of fitting the direct ringdown portion for initial data AE, UU, and UE, as well as for the reference binaries AB and UB. We use the same configuration as in Fig.~\ref{fig:mode_content}.  

Generally, we note that the extraction along the $z$ axis of the direct ringdown is more inaccurate for the configurations UU, UE, and UB. This is due to the fact that the merger is observed almost edge-on, and, especially for UU and UE, the duration of this ringdown is cut short by the arrival of the secondary image. Nevertheless, our results show that in general the dominant (quadrupolar) mode, and its amplitude, are extracted quite accurately in all cases. The subdominant mode seems to be compatible with the hexadecupolar $\ell=4$ mode at sufficiently late times. Its amplitude is recovered less accurately, with some large uncertainties in some cases. Moreover, in some of the scenarios considered here, we find certain evidence for a quadratic QNM with frequency $2\omega_{2,0}$, at relatively early ringdown start times. This mode is expected to be excited, but we believe neither our numerical simulations, waveform extraction procedure, and data analysis pipeline, are accurate enough to resolve it appropriately. A more thorough analysis of the excitation of nonlinear QNMs in the presence of companions is a worthwhile pursuit, that we leave open for the moment. 

Additionally, we report here the results obtained for the frequencies of the two free modes, that enter into the calculation of Fig.~\ref{fig:deltas}. Notice that the uncertainty reported here only estimates the fluctuations in the recovered frequencies due to varying the start time of the ringdown. This ignores other sources of error, both intrinsic (e.g. numerical error in the simulations), and systematic (the recovered frequencies would change slightly if we use $N=1$ or $N=3$ free modes. A Bayesian framework such as the one implemented in \texttt{BayRing}~\cite{Redondo-Yuste:2023seq, carullo_gregorio_2023_8284026} (see also~\cite{Dyer:2025iwj} for alternative approaches to this), would allow us to obtain yet more accurate results, with a better understanding of the uncertainty. 

\begin{table*}[h]
\begin{ruledtabular}
\begin{tabular}{llcccc}
  ID & Observer
     & $\Re[\omega_0]M_1$ & $-\Im[\omega_0]M_1$
     & $\Re[\omega_1]M_1$ & $-\Im[\omega_1]M_1$ \\
\hline
  \multirow{3}{*}{AE} & $(-100,\,0,\,10)$ & $0.3745(9)$ & $0.0875(6)$ & $0.799(3)$ & $0.097(3)$ \\
   & $(+100,\,0,\,10)$ & $0.3627(6)$ & $0.0873(4)$ & $0.770(5)$ & $0.098(6)$ \\
   & $(0,\,0,\,100)$ & $0.3700(4)$ & $0.0878(3)$ & $0.79(3)$ & $0.12(3)$ \\
\hline
  \multirow{3}{*}{AU} & $(-100,\,0,\,10)$ & $0.376(2)$ & $0.088(2)$ & $0.791(9)$ & $0.085(4)$ \\
   & $(+100,\,0,\,10)$ & $0.3572(3)$ & $0.0860(5)$ & $0.760(4)$ & $0.085(2)$ \\
   & $(0,\,0,\,100)$ & $0.3692(5)$ & $0.0877(4)$ & $0.78(3)$ & $0.12(3)$ \\
\hline
  \multirow{3}{*}{UU} & $(-100,\,0,\,0)$ & $0.3823(8)$ & $0.097(1)$ & $0.8(1)$ & $0.18(6)$ \\
   & $(+100,\,0,\,0)$ & $0.3588(3)$ & $0.0859(5)$ & $0.76(3)$ & $0.11(2)$ \\
   & $(0,\,0,\,100)$ & $0.3681(6)$ & $0.0930(5)$ & $0.770(4)$ & $0.089(2)$ \\
\hline
  \multirow{3}{*}{UE} & $(-100,\,0,\,0)$ & $0.3750(7)$ & $0.0881(5)$ & $0.79(3)$ & $0.11(3)$ \\
   & $(+100,\,0,\,0)$ & $0.3635(4)$ & $0.0871(5)$ & $0.77(2)$ & $0.11(2)$ \\
   & $(0,\,0,\,100)$ & $0.363(1)$ & $0.0949(9)$ & $0.806(3)$ & $0.08(2)$ \\
\end{tabular}
\end{ruledtabular}
\caption{%
  Frequencies inferred from $N=2$ free-mode fits of the first ringdown, at different observing points and for different initial data. We take the average and estimate the standard deviation from fits with ringdown start time $(t_0 - t_{\rm Rd_1})/M_1 = 10, 12.5, 15, 17.5,20$. The value in between brackets indicates the uncertainty in the last digit. 
  \label{tab:frequencies}
}
\end{table*}

Finally, we show the results of fitting the second ringdown with $N=2$ free damped sinusoids. The results of Table~\ref{tab:second_ringdown} are obtained from this analysis. 

\begin{figure*}
    \centering 
    \includegraphics[width=\linewidth]{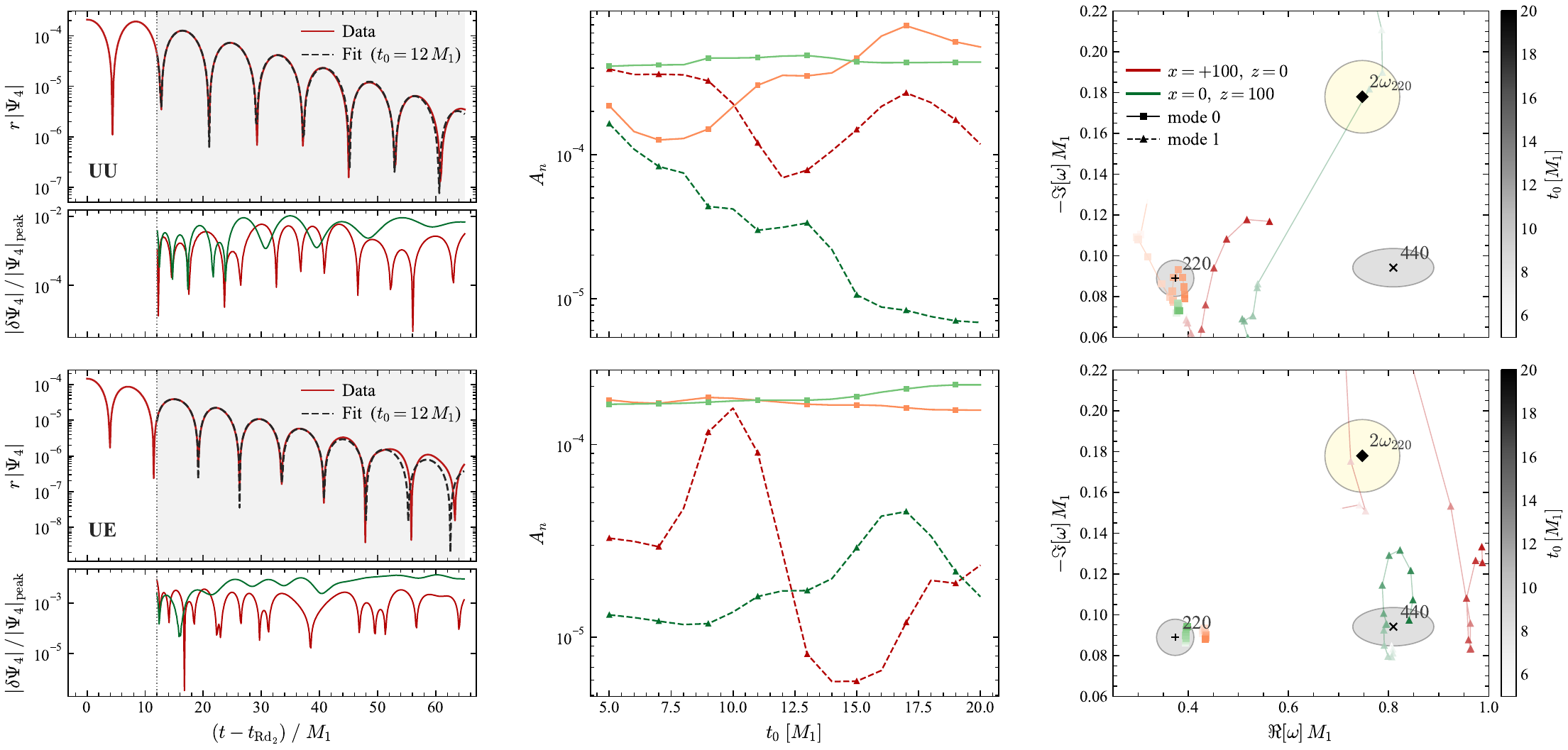}
    \caption{Same content as Fig.~\ref{fig:mode_content}, but for the second ringdown as observed for ID UU and UE, as seen from the top and the right of the merger. Notice the presence of two close-by modes from ID UU -- one of which we attribute to resonant mode excitation of the QNMs of the lens. We also highlight that the uncertainties in the extraction are larger than in e.g. Fig.~\ref{fig:modes_extra}, possibly hinting at more subtle distortions of the ringdown induced by lensing in the strong-field regime. }
    \label{fig:second_fits}
\end{figure*}
%%%%%%%%%%%%%%%%%%%%%%%%%%%%%%%%%%%%%%%%%%%%%%%%%%%
\section{Proof of time delay in geometric optics}\label{app:time_delay}
%%%%%%%%%%%%%%%%%%%%%%%%%%%%%%%%%%%%%%%%%%%%%%%%%%%
Here we prove Eq.~\eqref{eq:time_delay_1}. This is derived under a set of simplifying assumptions. \textit{(i)} The two BHs are stationary during the travel time of the null particle; \textit{(ii)} the null particle travels in a radial geodesic from one light ring to the other, so the problem is reduced to 1+1 dimensions; \textit{(iii)} the spacetime around both $M_1$ and $m_1$ can be described by the Schwarzschild metric with mass given by $M_1$ and $m_1$, respectively; \textit{(iv)} the individual metrics are glued at the Newtonian equipotential point, i.e., if the BHs are at a distance $d$, the gluing happens at a distance $d \, M_1/M$ from $M_1$, where $M=M_1+m_1$. In the Schwarzschild geometry with mass $M_1$ and standard $\{ t,r,\theta,\phi\}$ coordinates, a radial null geodescic has tangent vector 
\begin{equation}
    l^\mu = k (dt/dr, \pm 1, 0, 0) \, , \qquad k\in \mathbb R \, ,  \quad \frac{dt }{dr} = \left(1-\frac{2 M_1}{r}\right)^{-1} \, .
\end{equation}
Thus, the total time delay between the two light rings can be written as 
\begin{equation}
    \Delta t_{M_1\to m_1} = \int_{3M_1} ^{d \, M_1/M} \left(1-2M_1/r_1\right)^{-1}dr_1 + \int_{3m_1} ^{d \, m_1/M} \left(1-2m_1/r_2\right)^{-1}dr_2 \, .  
\end{equation}
We now simply change variables to $r_1 \to r=r_1 M/M_1$ and $r_2 \to r=r_2 M/m_1$. This immediately yields Eq.~\eqref{eq:time_delay_1}.

\end{document}